%% file: main-with-authos.tex
\documentclass{llncs}

\usepackage{algorithmicx}
\usepackage{algorithm}
\usepackage{algpseudocode}
\usepackage{amsfonts, amsmath, amssymb}
\usepackage{graphicx,color}
\usepackage{relsize}

\usepackage{hyperref}

\usepackage{ wasysym }
\usepackage{scalerel}

\usepackage{color,soul}
\usepackage{tikz}
\usepackage{multirow}

\usepackage{boldline}

\usetikzlibrary{arrows,shapes,positioning,calc,fit,backgrounds,intersections,through,automata}
\usetikzlibrary{decorations.pathreplacing,decorations.text}

\title{An Algebraic Framework for\\ Runtime Verification}
\author{Stefan Jak\v{s}i\'{c}\inst{1,2} \and Ezio Bartocci\inst{2} \and Radu Grosu\inst{2} \and Dejan Ni\v{c}kovi\'{c}\inst{1}}

\institute{AIT Austrian Institute of Technology GmbH, Vienna, Austria \and Vienna University of Technology, Vienna, Austria}

\begin{document}
\maketitle

\input ztmpops_macros
\input macros
\input abstract

\input intro
\input related

\input prelim

\input monitors

\input eval

\input conclusion

\bibliographystyle{plain}
\bibliography{ref}

\newpage

\appendix
\input specs
\input proof_th1

\end{document}

%% file: ztmpops_macros.tex



\newcommand\Dash{\scaleobj{0.9}{$\textendash$}}
\newcommand\Dia{\scaleobj{1.1}{\raisebox{-.4ex}
		{\rotatebox{45}{$\square$}}}}

\newcommand\DiaB{\scaleobj{1.2}{\raisebox{-.4ex}
		{\rotatebox{45}{$\square$}}}}

\newcommand\Boxx{\scaleobj{1.1}{\raisebox{-.1ex}{$\square$}}}
\newcommand\BoxxB{\scaleobj{1.3}{\raisebox{-.05ex}{$\square$}}}


\newcommand\cir{\scaleobj{1.3}{\ocircle}}

\newcommand\dia{\Dia}

\newcommand\boc{\BoxxB}

\newcommand\cirM{\scaleobj{1.3}{\circleddash}}

\newcommand\diaM{\mbox{\ \ooalign{$\raisebox{.35ex}\Dash$\cr\hidewidth\raisebox{-.0ex}
			{\hspace{+0.2pt}$\Dia$}\hidewidth\cr}\ }}

\newcommand\boxM{\mbox{\ \ooalign{$\raisebox{.35ex}\Dash$\cr\hidewidth\raisebox{-.0ex}
			{\hspace{+0.2pt}{$\BoxxB$}}\hidewidth\cr}\ }}

\newcommand\until{\hskip 1.5pt\hbox{$\cal U$}\hskip 0.5pt}
\renewcommand\unless{\hskip 1.5pt\hbox{$\cal W$}\hskip 0.5pt}
\newcommand\since{\hskip 1.5pt\hbox{$\cal S$}\hskip 0.5pt}
\newcommand\backto{\hskip 1.5pt\hbox{$\cal B$}\hskip 0.5pt}
\newcommand\ent{\hbox{\hskip 1.5pt\hbox{$=$\hskip
-3pt\hbox{$\succ$}\hskip 0.5pt}}}

\newcommand\Next{\cir}
\newcommand\X{\cir}
\newcommand\sU{\until}
\newcommand\wU{\unless}
\newcommand\sS{\since}
\newcommand\wS{\backto}
\newcommand\DiamondM{\diaM}
\newcommand\Previous{\cirM}
\newcommand\BoxM{\boxM}
\newcommand\G{\boc}
\newcommand\F{\dia}
\newcommand\opH{\boxM}
\newcommand\opP{\diaM}
\newcommand\XX{\Next}
\newcommand\Y{\Previous}
\newcommand\YY{\Previous}
\newcommand\ssU{\unless}
\newcommand\Sscr{\since}
\newcommand\fildia[1]{\opP}
\newcommand\filcir[1]{\Y}
\newcommand\filboc[1]{\opH}
\newcommand\wY{\filcir{\sim}}
\newcommand\emm[1]{\mbox{\em #1\/}}

%% file: macros.tex
\newcommand{\N}{\mathbb{N}}
\newcommand{\R}{\mathbb{R}}
\newcommand{\Q}{\mathbb{Q}}
\newcommand{\Z}{\mathbb{Z}}
\newcommand{\B}{\mathbb{B}}

\newcommand{\proj}[2]{\pi_{#1}(#2)}

\newcommand{\stl}{\textsc{STL}}
\newcommand{\mtl}{\textsc{MTL}}
\newcommand{\ltl}{\textsc{LTL}}

\newcommand{\algebra}{\mathcal{S}}


\newcommand{\ozero}{{\raisebox{0pt} {$e_{\oplus}$}}}
\newcommand{\oone}{{\raisebox{0pt} {$e_{\otimes}$}}}

\newcommand*{\mybox}[1]{\raisebox{3pt}{
  \framebox{\raisebox{-2.5pt}[0.15\baselineskip][0\baselineskip]{
    \hbox to 0.08cm{\hss#1\hss}}}}}
    
\newcommand*{\mylowbox}[1]{\raisebox{3pt}{
  \framebox{\raisebox{-2.5pt}[0.10\baselineskip][0\baselineskip]{
    \hbox to 0.08cm{\hss#1\hss}}}}}    
    
\newcommand{\metric}{\mathcal{M}}
\newcommand{\pterm}{\tau}

%

\newcommand{\boxzero}{{\raisebox{0pt} {$e^{\boxplus}$}}}
\newcommand{\boxone}{{\raisebox{0pt} {$e^{\boxtimes}$}}}

\newcommand{\lowboxzero}{{\raisebox{0pt} {$e^{\boxplus}$}}}
\newcommand{\lowboxone}{{\raisebox{0pt} {$e^{\boxtimes}$}}}

\newcommand{\ttrue}{\textsf{true}}
\newcommand{\ffalse}{\textsf{false}}

\newcommand{\domain}{D}
\newcommand{\vars}{X}
\newcommand{\predicate}{\psi}
\newcommand{\predicates}{\Psi}
\newcommand{\val}{v}

\newcommand{\aut}{\mathcal{A}}
\newcommand{\waut}{\mathcal{W}}
\newcommand{\weighted}{\waut}
\newcommand{\states}{Q}
\newcommand{\init}{I}
\newcommand{\final}{F}
\newcommand{\transitions}{\Delta}

\newcommand{\valnew}[1]{c'(#1)}
\newcommand{\valold}[1]{c(#1)}

\newcommand{\spec}{\varphi}
\newcommand{\trace}{\tau}
\newcommand{\traces}{\mathcal{T}}
\newcommand{\ppath}{\pi}
\newcommand{\pweight}{\mu}
\newcommand{\paths}{\Pi}
\newcommand{\tvalue}{\alpha}
\newcommand{\cost}{\lambda}

\newcommand{\vpd}{\textsf{vpd}}

\floatname{algorithm}{Algorithm}
\renewcommand{\algorithmicprocedure}{\textbf{Algorithm}}

\newcommand{\robustness}{\rho}


\newtheorem{nnthm}{\ \\Theorem}
\newtheorem{nnprop}{\ \\Proposition}
\newtheorem{nnlem}{\ \\Lemma}

\newtheorem{innercustomthm}{Theorem}
\newenvironment{customthm}[1]
  {\renewcommand\theinnercustomthm{#1}\innercustomthm}
  {\endinnercustomthm}

\newtheorem{innercustomlem}{Lemma}
\newenvironment{customlem}[1]
  {\renewcommand\theinnercustomlem{#1}\innercustomlem}
  {\endinnercustomlem}

\newcommand{\true}{\textsf{true}}
\newcommand{\false}{\textsf{false}}

%% file: abstract.tex
\begin{abstract}
Runtime verification (RV) is a pragmatic and scalable, yet rigorous
technique, to assess the correctness of complex systems,
\mbox{including} cyber-physical systems (CPS). By measuring how
robustly a CPS run satisfies a specification, RV allows in addition,
to quantify the resiliency of a CPS to perturbations. In this paper we
propose Algebraic \mbox{Runtime} \mbox{Verification} (ARV), a general,
semantic framework for RV, which takes \mbox{advantage} of the
monoidal~structure of~runs~(w.r.t.~\mbox{concatenation}) and the
semiring structure of a specification automaton (w.r.t.~choice and
concatenation), to compute in an incremental and \mbox{application}
specific fashion the resiliency measure. This allows us to
\mbox{expose} the core aspects of RV, by developing an abstract
monitoring \mbox{algorithm}, and to strengthen and unify the various
qualitative and quantitative approaches to RV, by instantiating choice
and concatenation with real-valued functions as dictated by the
application. We demonstrate the power and effectiveness of our
framework on two case studies from the automotive domain.
\end{abstract}

%% file: intro.tex
\section{Introduction}
\label{sec:intro}

Verification of realistic cyber-physical systems is still a challenge, as CPS exhibit both discrete and continuous
dynamics, software and hardware \mbox{components}, complex communication between sub-systems, and sophisticated
\mbox{interactions} with the physical environment. Runtime verification provides a pragmatic, yet rigorous solution, for
assessing CPS correctness at runtime.

RV for CPS~\cite{chapter5}, centered around Signal Temporal Logic (STL)~\cite{mn04}, has recently achieved noticeable
success. STL extends Metric Temporal Logic (MTL)\\\cite{koymans} with predicates over real-valued variables. The
monitoring algorithms for STL were first developed in~\cite{mn04,mn13} and implemented in a tool~\cite{amt}, giving thus
STL a practical purpose in the analysis of realistic CPS.

In a CPS context, RV measures the satisfaction robustness of a CPS run with respect to a
specification~\cite{fals2,fals1,staliro_base}. Most measures were associated to STL
specifications~\cite{fainekos-robust,robust1,robust2,fals1,fals2,staliro_base,breach,stlce,stltest} and to its variants
or extensions~\cite{stlstarq,avstl}. The general approach was to first develop the measure, based on the STL syntax, and
then use the measure for monitoring. This resulted in a plethora of measures, assessing
space~\cite{fainekos-robust,robust1}, time~\cite{robust1}, and averaging~\cite{avstl} robustness, respectively. A
discrete-time, weighted, edit-distance measure was proposed in~\cite{stledit}, and a space-robustness measure for Signal
Regular Expressions (SRE) in~\cite{OdedSRE}.

The proliferation of measures reflects the various needs of different applications, and the limitations of STL-based
measure definitions.
We propose Algebraic \mbox{Runtime} \mbox{Verification} (ARV), as a general, semantic
approach to measuring a run, with respect to a specification automaton. The main idea is to use the monoidal structure
of a run (with respect to concatenation), and the semiring structure of a specification (with respect to choice and
concatenation), to defining a measure in an incremental and application specific way, by allowing to associate arbitrary
real-valued functions to choice and concatenation.

ARV simplifies and unifies qualitative and quantitative approaches to the RV of CPS. It exposes the core of the RV
problem, and its underlying structure, allowing us thus to develop an abstract monitoring algorithm, that we instantiate
with different specification languages, as well as qualitative and quantitative semantics. The use of automata, admits
semantic and precise monitoring procedures, that are invariant to different syntactic representations of the same
specification. In addition, it opens the door for flexible code generation which can directly translate the abstract
monitoring procedure into real-time monitors, implemented in either software, embedded software, or hardware.

ARV starts from a regular specification formalism that admits an effective translation into symbolic
automata. It then decorates the symbolic acceptor with weights and associates a semiring to the resulting symbolic weighted
automaton (SWA). The weights in the SWA measure the distance at a given point in time between the current trace
observation and the constraints induced by the specification. ARV defines the monitoring procedure over the SWA as a
dynamic programming algorithm that computes the shortest path induced by an input trace.  By instantiating the semiring,
it provides various qualitative and quantitative semantics to the monitoring procedure without changing the underlying
algorithms.  We study the basic properties of our generic ARV framework and evaluate it with two case studies
from the automotive domain.

The rest of the paper is organized as follows. In Section 2 we discuss related work. In Section 3 we introduce the
theoretical background for formalizing ARV in Section 4. Section 5 includes two case studies performed on scientific benchmarks, as well as a precision comparison between ARV and tools that implement syntactic-based robustness degree. In Section 6 we draw
conclusions and discuss numerous possibilities for future work. Theorem proofs and STL and SRE semantics
are given in the appendices.
\vspace{-16pt}

%% file: related.tex
\section{Related Work}
\label{sec:related}

\vspace{-10pt}
The theoretical and practical concerns regarding symbolic automata and transducers are studied in~\cite{loris1,loris2,veanes_symbolic}. 
In our work, we use the theory developed for symbolic automata to develop our ARV framework.

An algebraic framework for the basic properties of the weighted automata is studied in~\cite{edit-weighted}.  
The shortest-distance problem in weighted automata is investigated in~\cite{mohri_semiring}. 
It generalizes the Bellman-Ford algorithm~\cite{bellman1958routing} to non-idempotent semirings. 
We are interested in a Hausdorff measure and hence restrict our attention to additively idempotent semirings.
In constrast to fixed edge weights, we use SWA with dynamic weights which depend on current trace valuation. 

Quantitative semantics for temporal logics based on the (spatial) infinite norm were studied in~\cite{rizk,fainekos-robust,robust1}. Spatial robustness monitoring 
is implemented in S-TaLiRo~\cite{staliro_base} and Breach~\cite{breach} tools. The spatial robustness was complemented with time robustness in~\cite{robust1} and 
with a combined time-space robustness based on $(\epsilon, \tau)$-similarity in~\cite{AbbasMF14}.
In~\cite{avstl}, the authors extend $\stl$ with {\em averaged} temporal operators. Determining robustness 
of hybrid systems using self-validated arithmetics is shown in~\cite{njofra}. The weighted Hamming and edit distances between behaviors are 
proposed in~\cite{roopsha_robustness13}, where the authors use it to develop procedures for reasoning about the Lipshitz-robustness of 
Mealy machines and string transducers. The authors of~\cite{LOLA} propose an online monitoring procedure where semantics describe the relation 
between input and output streams.  In~\cite{BartocciBLN17} the authors used an algebraic approach to define the
robustness for a spatio-temporal extension of STL considering only the MinMax semiring.  Their approach works 
at the syntax level of the specification, resulting to be less precise than the one proposed here. Similarly to our work,~\cite{rodionova2016temporal} explores different interpretation of temporal logic operators. In contrast to ARV, these interpretations are applied directly on the syntax and semantics of the logic, with the aim to demonstrate a relation between temporal logic operators and convolution. We also mention the work on quantitative languages~\cite{chatterjee2008quantitative} that is studied over infinite words and is complementary to our 
work.


The problem of online robustness monitoring was studied more recently in~\cite{DokhanchiHF14,DeshmukhDGJJS17}. 
The authors of~\cite{DokhanchiHF14} propose a predictor-based online monitoring approach, in contrast to our 
black-box view of monitoring. In~\cite{DeshmukhDGJJS17}, the authors propose an interval-based approach of 
online evaluation that allows estimating the minimum and the maximum robustness with respect to both the 
observed prefix and unobserved suffix of the trace. Instead, our 
robustness gives the distance of the observed prefix from the of the specification at every point in time.
\vspace{-24pt}

%% file: prelim.tex

\section{Background}
\label{sec:prelim}

\vspace{-8pt}
We first introduce the background needed to develop our algebraic runtime verification algorithm. In particular, 
we define semirings, metric spaces and distances, specification languages, and symbolic automata and SWA.
\vspace{-12pt}

\subsection{Semirings}

Semirings are one of the most important algebraic structures, laying the foundation to both continuous and discrete
mathematics. They help finding similarities between the two domains, even in places where these are not at all obvious.

\vspace{-8pt}
\begin{definition}[Semiring]
\label{def:semiring}
A {\em semiring} is the tuple $\algebra = (S, \oplus, \otimes, \ozero, \oone)$, where $S$ is a set equipped with two 
binary operations, {\em addition} ($\oplus$) and {\em multiplication} ($\otimes$), and two identity elements, $\ozero$ and $\oone$, such that: 
\vspace{-4pt}
\begin{itemize}
\item $(S, \oplus, \ozero)$ is a commutative monoid with identity element $\ozero$;
\item $(S, \otimes, \oone)$ is a monoid with identity element $\oone$;
\item $\otimes$ distributes over $\oplus$; and 
\item $\ozero$ is an annihilator element for $\otimes$.
\end{itemize}
\end{definition}

\vspace{-8pt}
We say that a semiring is {\em commutative} if the $\otimes$-multiplication operation is commutative. 
A semiring is said to be {\em additively (multiplicatively) idempotent} if for all $s \in S$, we have that $s \oplus s = s$ $(s \otimes s = s)$. 
We say that a semiring is {\em idempotent} if it is both additively and multiplicatively idempotent. We say that a semiring is {\em bounded} if 
$\oone$ is an annihilator element for $\oplus$.
We note that a bounded semiring is also additively idempotent~\cite{mohri_semiring}.

\vspace{-8pt}
\begin{example} 
\label{ex:one}
We depict in Table~\ref{tab:semiring} several examples of semiring structures that we use in this paper. We note that the Boolean and 
the MinMax semirings are both commutative and idempotent. The tropical semiring is commutative and additively idempotent. All three semirings are bounded. Note that  
we use a non-standard definition of the Boolean and MinMax semirings, in which $\oplus$ corresponds to $\wedge$ and $\min$, while $\otimes$ corresponds to $\vee$ and $\max$, 
respectively. 

\vspace{-24pt}

\begingroup
\begin{table}[htb]
\centering

$$
\begin{array}{||l|c|c|c|c|c||}
\hline
~\textrm{Semiring}~			& ~S~					&~\oplus~		& ~\otimes~		& ~\ozero~		& ~\oone~
\\ 
\hline
~\textrm{Boolean}~			& \{ 0, 1 \}				& \wedge			& \vee		& 1~				& ~0~ 				\\
~\textrm{MinMax}~			& ~\R_{+} \cup \{ \infty \} ~ & ~ \min ~ & ~ \max ~& ~\infty~			& ~0~ 				\\
~\textrm{Tropical}~			& ~\R_{+} \cup \{ \infty \} ~ & ~ \min	~ & ~+~& ~\infty~			& ~0~ 				\\
\hline
\end{array}
$$
\setlength{\abovecaptionskip}{-2pt}
\setlength{\belowcaptionskip}{-6pt}
\caption{Examples of semirings.}
\label{tab:semiring}
\end{table}
\endgroup

\end{example}

\vspace{-24pt}

%
%
%
%
Additively idempotent semirings defined over sets of Booleans, naturals and reals admit a natural order between the elements in the set.

\vspace{-4pt}
\begin{definition}[Natural order on $S$]
\label{def:order}
Let $(S, \oplus, \otimes, \ozero, \oone)$ be an additively idempotent semiring. We define the {\em natural order} on $S$ as the relation $\sqsubseteq$:$~
(a \sqsubseteq b) \\ \leftrightarrow (a \oplus b = a).
$
\noindent If $(S, \oplus, \otimes, \ozero, \oone)$ is also multiplicatively idempotent, we additionaly require:
$
(a \sqsubseteq b)~\leftrightarrow~(a \otimes b = b).
$
\end{definition}

\vspace{-4pt}
\begin{lemma}[\cite{mohri_semiring}]
\label{lemma:order}
Let $(S, \oplus, \otimes, \ozero, \oone)$ be an additively idempotent semiring. The natural order $\sqsubseteq$ on $S$ defines a partial order.
\end{lemma}

\vspace{-4pt}
We now define the monotonicity of semirings, an important property that will allow us factoring and thus simplifying our RV operations.

\vspace{-4pt}
\begin{definition}[Negative and monotonic semirings]
\label{def:mono}
Let $(S, \oplus, \otimes, \ozero, \oone)$ be a semiring. We say that $S$ is {\em negative} if $\oone \sqsubseteq \ozero$. We say that $S$ is 
{\em monotonic} if for all $a,b,c \in S$:
\vspace{-4pt}
\begin{enumerate}
\item $a \sqsubseteq b \rightarrow (a \oplus c) \sqsubseteq (b \oplus c)$
\item $a \sqsubseteq b \rightarrow (a \otimes c) \sqsubseteq (b \otimes c)$
\item $a \sqsubseteq b \rightarrow (c \otimes a) \sqsubseteq (c \otimes b)$
\end{enumerate}
\end{definition}

\vspace{-4pt}
\begin{lemma}[\cite{mohri_semiring}]
\label{lemma:mono}
Let $\algebra$ be an additively idempotent semiring $(S, \oplus, \otimes, \ozero, \oone)$ equipped with the natural order $\sqsubseteq$ on $S$. Then $S$ is 
both negative and monotonic.
\end{lemma}

\vspace{-4pt}
\begin{lemma}
\label{lemma:one-min}
Let $\algebra = (S, \oplus, \otimes, \ozero, \oone)$ be an additively idempotent, negative and monotonic semiring. Then, for all 
$a \in S$, $\oone \sqsubseteq a \sqsubseteq \ozero$.
\end{lemma}

\vspace{-4pt}

\subsection{Metric Spaces and Distances}

A metric space is a set $\metric$ possesing a distance among its elements. The distance $d(m_{1}, m_{2})$ between two elements $m_{1}, m_{2} \in \metric$ is a positive real value in $\R_+$.  

\vspace{-8pt}
\begin{definition}[Metric space and distance]
\label{def:distance}
Given a set $S$, let $d~:~\metric \times \metric \rightarrow \R_+$ be a {\em distance}. 
Then $\metric$ is a  {\em metric space} with the {\em distance measure} $d$, if:
\vspace{-4pt}
\begin{enumerate}
\item $d(m_{1},m_{2}) \geq 0$ for all $m_{1},m_{2}$ in $\mathcal{M}$; 
\item $d(m_{1},m_{2}) = 0$ if and only if $m_{1} = m_{2}$;
\item $d(m_{1},m_{2}) = d(m_{2},m_{1})$ for all $m_{1},m_{2}$ in $\mathcal{M}$; and 
\item $d(m_{1},m_{2}) \leq d(m_{1},m) + d(m,m_{2})$ for all $m,m_{1},m_{2}$ in $\mathcal{M}$.
\end{enumerate}
\end{definition}

\vspace{-4pt}
Since we reason about real-valued behaviors and the distances between them, we 
are interested in semirings defined over (subsets of) reals, see Example~\ref{ex:one}.
Given $m \in \mathcal{M}$ and $M \subseteq \mathcal{M}$, we can lift the above definition to 
reason about the distance\footnote{Since $d(m,M)$ is comparing an element to a set, strictly speaking it is not a distance.} 
between an element $m$ of $\mathcal{M}$  and the subset $M$ of $\mathcal{M}$ to define a Hausdorff-like measure. We use the $\oplus$-addition to 
combine individual distances between $m$ and the elements in $M$ and fix $\oone$ to $0$. We also need a special value when we compare $m$ to an empty set 
and define $d(m, \emptyset) = \ozero$.

\vspace{-4pt}
$$
d(m,M) = \left \{
				\begin{array}{ll}
				\ozero & \textrm{if }  M \textrm{ is empty} \\
				\oplus_{m' \in M} d(m,m') & \textrm{otherwise}
				\end{array}
				\right.
$$
\vspace{-4pt}
We define the {\em robustness degree} $\robustness(m,M)$ of $m$ w.r.t. the set $M$ as follows:
$$
\robustness(m,M) = 	\left \{
				\begin{array}{ll}
				d(m,\mathcal{M} \backslash M) & \textrm{if } m \in M \\
				-d(m,M) & \textrm{otherwise}
				\end{array}
				\right.
$$

\vspace{-24pt}
\subsection{Traces and Specification Languages}
\label{sec:trace}

\vspace{-4pt}
Let $\vars$ denote a set of variables defined over a {\em domain} $\domain$. 
We denote by $\val:\vars \rightarrow \domain$ the {\em valuation} function that maps a variable in 
$\vars$ to a value in $\domain$.
We denote by $\trace = \val_{1}, \ldots, \val_{n}$ a {\em trace} over $\vars$ and by $\traces(\vars)$ 
the set of all traces over $\vars$. 

A specification $\spec$ over a set of variables $\vars$, regardless of the formalism used, defines a {\em language} $L(\varphi) \subseteq \traces(\vars)$ 
that partitions the set of all traces over $\vars$. 

\vspace{-4pt}
\begin{definition}[Trace-specification distance]
\label{def:tr-spec}
Let $v$ and $v'$ be two valuations over $\domain^{|X|}$, $\trace$ and $\trace'$ two traces over $X$ of size $m$ and $n$  and $\spec$ a specification over $X$. We then 
have:
\vspace{-8pt}
$$
\begin{array}{lcl}
d(v,v') & = & \otimes_{x \in X} d(v(x), v'(x)) \\
d(\trace, \trace') & = & \left \{
				\begin{array}{ll}
				\ozero & \textrm{if }  m \neq n \\
				\otimes_{1 \leq i \leq m} d(v_{i},v'_{i}) & \textrm{otherwise}
				\end{array}
				\right. \\
d(\trace, \spec) & = & \oplus_{\trace' \models \spec} d(\trace, \trace') \\
\end{array}
$$

$
$.
\end{definition}

\vspace{-16pt}
In this paper, we consider specification languages defined over discrete time and real-valued variables that 
are {\em regular}. Examples of specification languages that fall into this category are Signal Temporal Logic (STL) and Signal Regular Expressions (SRE), both interpreted 
over discrete time. We briefly recall the syntax of STL and SRE and refer to their semantics in Appendix~\ref{sec:spec}.

We consider $\stl$ with both {\em past} and {\em future} operators interpreted over 
digital signals of final length.
We assume that $\domain$ is a metric space equipped with a distance $d$.
The syntax of a $\stl$ formula $\varphi$ over $X$ is defined by the grammar\footnote{STL interpreted over discrete time is equivalent to LTL extended with 
predicated over real-valued variables.}:
\vspace{-4pt}
$$
\begin{array}{lcl}
\varphi & := & x \sim u~|~\neg \varphi~|~\varphi_1 \vee \varphi_2~|~\varphi_1 \until_I \varphi_2~|~\varphi_1 \since_I \varphi_2 \\
\end{array}
\vspace{-4pt}
$$
\noindent where $x \in X$, $\sim \in \{ <, \leq \}$, $u \in \mathbb{D}$, 
$I$ is of the form $[a,b]$ or $[a, \infty)$ such that $a,b \in \mathbb{N}$ and 
$0 \leq a \leq b$. The other standard (Boolean and temporal) operators are derived from 
the basic ones. For instance, {\em eventually} $\F_I \varphi$ is defined as $\top \until_I \varphi$, 
while {\em always} $\G_I \varphi$ corresponds to $\neg \F_I \neg \varphi$.

The syntax of a SRE formula $\varphi$ over $X$ is defined by the grammar:
\vspace{-4pt}
 $$
\begin{array}{lcl}
b & := & x \sim u~|~\neg b~|~b_{1} \vee b_{2} \\
\varphi & := & \epsilon~|~b~|~\varphi_{1} \cdot \varphi_{2}~|~\varphi_{1} \cup \varphi_{2}~|~\varphi_1 \cap \varphi_2~|~\varphi^{*}~|~\langle \varphi \rangle_{I} \\
\end{array}
$$

\noindent where $x \in X$, $\sim \in \{ <, \leq \}$, $u \in \mathbb{D}$, 
$I$ is of the form $[a,b]$ or $[a, \infty)$ such that $a,b \in \mathbb{N}$ and 
$0 \leq a \leq b$. Although we interpret SRE over discrete time, we interpret its operators following the style of continuous time TRE. As a consequence, 
a signal segment that matches a predicate such as $x \leq 5$ means that it matches it for a strictly positive duration. The time duration operator $\langle \varphi \rangle_{I}$ 
is matched by a segment if it has a duration in $I$.

\vspace{-4pt}
\begin{example}
\label{ex:spec}
Consider the requirement `'There must be a point in time within the trace where (1) $x$ is smaller or equal than $3$, and (2) both $x$ is smaller or equal than $5$ and $y$ is
greater or equal than $6$ for the duration of at least two time steps. We formalize the above requirement as the STL specification 
$\varphi_{1} \equiv \F(x \leq 3 \wedge \G_{[0,1]} (x \leq 5 \wedge y \geq 6))$ and the SRE specification 
$\varphi_{2} \equiv \top \cdot ((x \leq 3) \cdot \top) \cap \langle x \leq 5 \wedge y \geq 6 \rangle_{[1,1]}) \cdot \top$. 
\end{example}
\vspace{-20pt}
\subsection{Symbolic and Symbolic Weighted Automata}
\label{sec:aut}

Let $\domain = \R$ be the domain of reals and $\vars$ the set of variables defined over $\domain$. 
%
%
We now define predicates over variables in $\vars$.
\vspace{-4pt}
\begin{definition}[Predicate]
\label{def:pred}
We define the syntax of a {\em predicate} $\predicate$ over $\vars$ with the following grammar:
$
\begin{array}{lll}
\predicate & := & \bot~|~\top~|~x \preceq k~|~\neg \psi~|~\psi_{1} \vee \psi_{2}~|~\psi_{1} \wedge \psi_{2} \\
\end{array}
$, where $\preceq \in \{<, \leq\}$, $x \in \vars$ and $k \in \domain$. 
\end{definition}

\vspace{-4pt}
We denote by $\predicates(\vars)$ the set of all predicates over $\vars$.
We say that a valuation $\val$ {\em models} 
the predicate $\predicate$, denoted by $\val \models \predicate$, iff $\predicate$ evaluates to true under $\val$.
We note that $x \preceq k$ plays the role of basic propositions. In our framework, the predicates come from 
specifications, hence we allow predicates in arbitrary form. In order to define the subsequent RV algorithms, we 
need to transform arbitrary predicates into {\em (minimal) disjunctive normal form (DNF)}.

\vspace{-4pt}
\begin{definition}[Predicate in Disjunctive Normal Form (DNF)]  
\label{def:DNF}
A predicate in disjunctive normal form is generated by the following grammar:\\
\begin{minipage}{7em}
\footnotesize
$\psi  :=  \psi^{c}~|~\psi^{c} \vee \psi$
\end{minipage}\hspace{4ex}
\begin{minipage}{8em}
\footnotesize
$\psi^{c} :=  \psi^{l}~|~\psi^{l} \wedge \psi^{c} $
\end{minipage}\hspace{4ex}
\begin{minipage}{13em}
\footnotesize
$\psi^{l}  :=  \top~|~\bot~|~(x \preceq k)~|~\neg (x \preceq k)$
\end{minipage}

\noindent where $\preceq \in \{<, \leq\}$, $x \in \vars$ and $k \in \domain$. 
The predicate has the following structure:
 \vspace{-4pt}
$$ \psi = \bigvee^{h}_{i=1} \psi^{c}_{i},\;\; \psi^{c}_{i} = \bigwedge^{m(i)}_{h=1}  \psi^{l}_{i,h}$$ where $h$ is the 
number of clauses, each clause $i$ is a conjunction of $m(i)$ literals and 
each literal can be either a basic proposition, its negation, true or false.  
\end{definition}

\begin{definition}[Predicate in $\wedge$-minimal DNF]
\label{def:minDNF}
A predicate $\predicate$ is expressed in a 
$\wedge$-minimal DNF if it satisfies the following properties:
$$   \bigwedge_{i=1}^{h} \bigwedge_{s=1,r = 1, s \neq r}^{m(i)} ( \psi^{l}_{i,s} \rightarrow \psi^{l}_{i,r})  = \bot  $$

\end{definition}

\begin{example}
The predicate $\psi_{1} \equiv (x \leq 3 \wedge x \leq 5 \wedge y \leq 5) \vee (z > 0)$ is in DNF, while the predicate 
$\psi_{2} \equiv (x \leq 3 \wedge y \leq 5) \vee (z >0)$ is in $\wedge$-minimal DNF.
\end{example}
 




We lift the definition of a distance between two valuations to the distance between a valuation and a predicate by $\oplus$-summing the distances between the  
valuation and the set of valuations defined by a predicate.

\begin{definition}[Valuation-predicate distance]
\label{def:pred-val}
Given a valuation $v \in D^{|\vars|}$ and a predicate $\psi \in \predicates (\vars)$, we have that:
$
d(v, \psi) = \mathlarger{\oplus}_{v' \models \psi}  \mathlarger{\otimes}_{x \in \vars} ~d(v(x), v'(x))
$.
\end{definition}

This completes our definitions for computing the distance between an observation and a specification at a 
single point in time. We now concentrate on the dynamic (temporal) aspect of the specification.
%
%
We first define {\em symbolic} and {\em symbolic weighted automata}.

\begin{definition}[Symbolic and Symbolic Weighted Automata]
\label{def:wsa}
A {\em symbolic automaton} (SA) $\aut$ is the tuple $\waut = (\vars, \states, \init, \final, \transitions)$, where $\vars$ is a finite set of variables defined 
over a domain $\domain$, $\states$ is a final set of {\em locations}, $\init \subseteq \states$ is the set of {\em initial} states, $\final \subseteq \states$ is 
the set of {\em final} states and $\transitions \subseteq \states \times \predicates(\vars) \times \states$ is the {\em transition relation}. A {\em symbolic weighted automaton} $\waut$ is the pair $(\aut, \cost)$, where $\aut$ is a symbolic automaton and  
$\cost~:~\transitions \times \domain^{|\vars|} \rightarrow \domain$ is the {\em weight} function. 
\end{definition}

In the following, we assume that the weights are elements of the semiring $(S, \oplus, \otimes, \ozero, \oone)$. We use $\otimes$ to compute 
the weight of a path by $\otimes$-multiplying the weights of the transitions taken along that path. We use $\oplus$ to $\oplus$-sum the 
path weights induced by an input trace. 
We now formalize the above notions. 

A {\em path} $\ppath$ in $\aut$ is a finite alternating sequence of locations and transitions 
$\ppath = q_{0}, \delta_{1}, q_{1}, \ldots, q_{n-1}, \delta_{n}, q_{n}$ such that $q_{0} \in \init$ and for all $1 \leq i \leq n$, $(q_{i-1}, \delta_{i}, q_{i}) \in \transitions$. 
We say that the path $\pi$ is {\em accepting} if $q_{n} \in \final$. We say that that a trace $\trace = \val_{1}, \val_{2}, \ldots, \val_{n}$ {\em induces} a path 
$\ppath = q_{0}, \delta_{1}, q_{1}, \ldots, q_{n-1}, \delta_{n}, q_{n}$ in $\aut$ if for all $1 \leq i \leq n$, $v_{i} \models \predicate_{i}$, where 
$\delta_{i} = (q_{i-1}, \predicate_{i}, q_{i})$. We denote by 
$\paths(\trace) = \{ \ppath~|~\ppath \in \final \; \textrm{and} \; \trace \; \textrm{induces} \; \ppath \; \textrm{in} \; \aut \}$ the 
set of all accepting paths in $\aut$ induced by trace $\trace$. 

\begin{definition}[Path and trace value]
\label{def:path_weight}
The {\em path value} $\pweight(\trace, \ppath)$ of a path $\ppath$ induced in $\aut$ by a trace $\tau$ is defined as:
$~ 
\pweight(\trace, \ppath) = \mathlarger{\otimes}_{1 \leq i \leq n} \lambda(\val_{i}, \delta_{i}).
$
The {\em trace value} $\tvalue(\trace, \waut)$ of a trace $\trace$ in $\waut$  is defined as:
$~
\tvalue(\trace, \waut) = \mathlarger{\oplus}_{\ppath \in \paths(\trace)} \pweight(\trace, \ppath).~
$
\end{definition}

%% file: monitors.tex

\vspace{-16pt}
\section{Algebraic Monitors for Correctness and Robustness}
\label{sec:mon}

\begin{figure}
\centering
\scalebox{0.75}{ 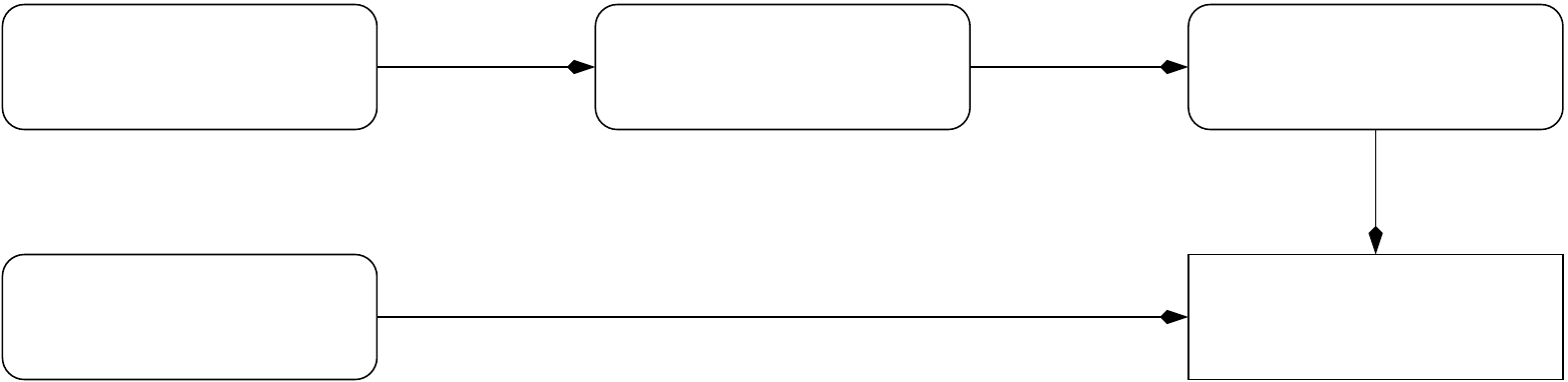 }
\setlength{\abovecaptionskip}{-4pt}
\caption{Computation of $d(\trace,\spec)$.}
\label{fig:steps}
\end{figure}

In this section, we develop ARV, a novel procedure for abstract computation of the {\em robustness degree} of a discrete signal with respect 
to a specification $\spec$.  The proposed methods consists of several steps, illustrated in Figure~\ref{fig:steps}. We 
first translate the specification $\spec$ into a symbolic automaton $\aut_{\spec}$ that accepts the same language as the specification. 
The automaton $\aut_{\spec}$ treats timing constraints from the formula in an enumerative fashion, but keeps symbolic 
guards on data variables. We then decorate $\aut_{\varphi}$ with weights on transitions, thus obtaining the {\em symbolic weighted automaton} $\waut_{\spec}$.
We propose an abstract algorithm for computing the distance between a trace $\tau$ and a specification $\spec$ by reducing it to the 
problem of finding the shortest path in $\waut_{\spec}$ induced by $\trace$.  
Computing the robustness degree between the trace $\trace$ and the specification $\spec$ follows from combining  
the computed distance from the specification $\spec$ and its negation $\neg \spec$. 

\vspace{-12pt}

\vspace{-24pt}
\subsection{From Specification to Symbolic Weighted Automaton}
\label{sec:translation}

We assume in this paper an effective translation \textsf{T} of a regular specification language $\spec$ to symbolic automata $\aut_{\spec}$ that accepts 
the language of $\spec$. 
For STL and SRE defined over discrete time, such translation is a moderate adaptation of standard 
methods, including on-the-fly tableau construction~\cite{ltl-tableau} and temporal testers~\cite{testers}. 
During the translation, we decorate the transitions in the  symbolic automaton with weight functions that measure the 
distance between observed valuations  and the predicate $\predicate$ associated to the transition, i.e. we set 
instantiate the weight function $\cost$ to $\cost(\delta, v) = d(v, \predicate)$, for all $\delta = (q, \predicate, q') \in \Delta$.

\vspace{-4pt}
\begin{example}
We illustrate this step on specifications $\varphi_{1}$ and $\varphi_{2}$ from Example~\ref{ex:spec}:

  
	\begin{tabular}{lll}
\centering
		$ \varphi_1$ & $\equiv$ & $\F ( x \leq 5~\wedge~\G_{[0,1]}(x \leq 3~\wedge~y > 6) )$ \\
		$ \varphi_2$ & $\equiv$ & $\top~\cdot~( ( x \leq 5 \cdot \top )~\cap~\big \langle x \leq 3 \wedge y \geq 6 \big \rangle_{[1,1]})~\cdot~\top~ $\\
	\end{tabular}

\noindent
In Figure~\ref{fig:exampleWSA} we depict the SWA that accepts the language of $\varphi_{1}$ and $\varphi_{2}$. The $\wedge$-minimal predicates are shown in the boxes above the transitions.

\vspace{-16pt}
\begin{figure}
\label{fig:one}
\centering
\hspace*{-0.35cm}\scalebox{0.9}{\input{example}}
\vspace*{-15mm}
\setlength{\abovecaptionskip}{-6pt}
\caption{SWA $\waut$ that accepts the language of $\varphi_1$ and $\varphi_2$.}
\label{fig:exampleWSA}
\end{figure}
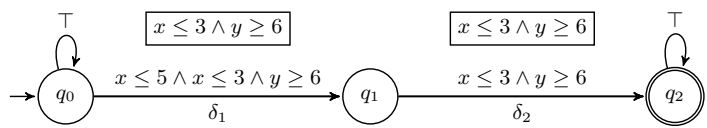

\label{ex:exampleWSA}
\end{example}

\vspace{-8pt}
\subsection{Valuation-Predicate Distance Computation}
\label{sec:distance}

We propose an effective procedure for computing the distance between a valuation $v$ and a predicate $\predicate$. 
The procedure is shown in Algorithm~\ref{alg:pred-val} and works as follows. The input to the procedure 
is a valuation $v$ and a predicate $\psi$ in DNF. The computation of the distance between $v$ and $\predicate$ is done inductively in 
the structure of $\predicate$. In the base case when $\predicate$ is an atomic predicate, the distance between $v$ and $\predicate$ is 
$\oone$ if $v$ satisfies $\predicate$, and it is equal to $d(v(x), k)$ otherwise. We compute logic $\vee$ and 
$\wedge$ by interpreting them as $\oplus$-addition and $\otimes$-multiplication.

\vspace{-16pt}

\noindent
\begin{minipage}[t]{0.47\textwidth}
\begin{flushleft}

\begin{algorithm}[H]
\small
  \caption{\vpd$(v, \predicate)$}
\begin{algorithmic}[1]

\Require Predicate $\predicate \equiv \bigvee_{i} \bigwedge_{j} p_{ij}$


\If {($\predicate$ is UNSAT)}
    \Return  \ozero
\EndIf	
\If {($\predicate = x \preceq k$) or ($\predicate =\neg (x \preceq k))$}
    
    \If {($\val_{(x)} \models \psi$)}
    \Return  \oone
    \Else \textbf{~return} $d(v(x), k)$
    \EndIf

\ElsIf {($\predicate =\predicate_{1} \vee \predicate_{2} $)}
	\State \textbf{return} $\textsf{vpd}(\val, \predicate_{1}) \oplus \textsf{vpd}(\val, \predicate_{2})$

\ElsIf {($\predicate =\predicate_{1} \wedge \predicate_{2} $)}
	\State \textbf{return} $\textsf{vpd}(\val, \predicate_{1}) \otimes \textsf{vpd}(\val, \predicate_{2})$
\EndIf
\end{algorithmic}
\label{alg:pred-val}
\end{algorithm}
\end{flushleft}

\end{minipage}
~~
\begin{minipage}[t]{0.49\textwidth}
\begin{algorithm}[H]
\small
  \caption{$\wedge\textsf{-min}(v, \predicate)$}
  \begin{algorithmic}[1]
\Require Predicate $\predicate \equiv \bigvee_{i} \bigwedge_{j} p_{ij}$

\State Let $P_{i} = \{ p_{i1}, \ldots, p_{ij} \}$
\State Let $\mathcal{P} = \{ P_{1}, \ldots, P_{i} \}$

\ForAll{$P \in \mathcal{P}$}
	\ForAll{$p_{1}, p_{2} \in P$ s.t. $p_{1} \neq p_{2}$}
		\If{$p_{1} \rightarrow p_{2}$}
			\State $P \leftarrow P \backslash \{ p_{2} \}$
		\EndIf
	\EndFor
\EndFor

\State \textbf{return} $\bigvee_{P \in \mathcal{P}} \bigwedge_{p \in P} p$
\end{algorithmic}

\label{alg:wedge-min}
\end{algorithm}
\end{minipage}

\vspace{2pt}
The procedure presented in Algorithm~\ref{alg:pred-val} indeed computes the distance from Definition~\ref{def:pred-val} for (1) idempotent 
semirings and (2) additively idempotent semirings with the predicate given in $\wedge$-minimal DNF. The transformation of arbitrary predicates 
to DNF is standard, while we present the $\wedge$-minimization of a predicate in DNF in Algorithm~\ref{alg:wedge-min}. In the following example, 
we give intuition why predicates must be in $\wedge$-minimal DNF if the semiring is only $\oplus$-idempotent.

\vspace{-4pt}
\begin{example}
Consider the term $x \leq 3 \wedge x \leq 5$ from the transition $\delta_{1}$ in Figure~\ref{fig:exampleWSA} that defines the set of 
valuations $\{v(x)~|~v(x) \leq 3\}$, its semantically equivalent $\wedge$-minimal representation $x \leq 3$ and the valuation $v(x) = 6$. It is clear that 
$d(v, x \leq 3 \wedge x \leq 5) = d(v, x \leq 3) = 3$. Let us consider the tropical semiring and the computation of $\vpd$. We illustrate 
the need for the $\wedge$-minimal predicate in Figure~\ref{fig:idem}. We have that 
$\vpd(v, x\leq 3) = 3$, but $\vpd(v, x \leq 3 \wedge x \leq 5) = 1 \otimes 3 = 1 + 3 = 4$. Due to the non-minimality of the term and the 
additive nature of $+$, we incorrectly accumulate the distance from $v(x)$ to the atomic predicate $x \leq 5$.

\vspace{-12pt}

\begin{figure}
\label{fig:idem}
\centering
\hspace*{-0.35cm}\scalebox{0.75}{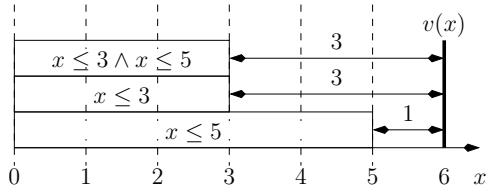 }
\setlength{\abovecaptionskip}{-4pt}
\caption{Example of a distance between $v$ and $\psi$.}
\label{fig:idem}
\end{figure}

\end{example}


\vspace{-24pt}
\begin{theorem}
\label{thm:valpred}
Given a predicate $\predicate$ in DNF, a valuation $v$ and the distance 
$d(v, \predicate)$ defined over a bounded semiring $S$, we have that $\textsf{vpd}(v,\predicate) = d(v, \predicate)$ 
if: (1) $S$ is idempotent; or (2) $\predicate$ is in $\wedge$-minimal DNF.
\end{theorem}



\vspace{-16pt}
\subsection{Trace Value Computation}
\label{sec:trval}

In this section, we present a dynamic programming procedure (see Algorithm~\ref{alg:val}) for computing the value of a trace $\tau = v_{1}, \ldots, v_{n}$ in 
a symbolic weighted automaton $\waut$. We assume that the weights are defined over a semiring $S$. The algorithm first assigns to every state $q$ the initial cost, depending whether $q$ is initial or not. At every step $i \in [1,n]$ and for every state $q$ of $\waut$, the procedure 
computes the cost of reaching $q$ with the $i$-prefix of $\tau$. The procedure uses the $\otimes$-multiplication to aggregate the valuation-predicate distances 
collected along every path $\pi$ induced by $\tau$ and thus compute the path weight and the $\oplus$-addition to combine the weights of all the accepting paths induced by $\tau$. 

\vspace{-24pt}

\noindent
\begin{minipage}[t]{0.58\textwidth}
\begin{algorithm}[H]
\small

  \caption{\textsf{val}$(\tau, \waut)$}
\begin{algorithmic}[1]
\ForAll{$q \in Q$}
  \State $\valold{q,0} \gets (q \in I)~?~\oone~:~\ozero$ 
\EndFor

\For{$i = 1~to~n$}
\State $\valold{q,i} \gets \oplus_{(s, \psi, q) \in \Delta} (\valold{s,i-1} \otimes \vpd(v_{i}, \psi))$
\EndFor


  
  

\State \textbf{return} $\mathlarger{\oplus}_{q \in F}~ \valold{q,n}$
\end{algorithmic}
\label{alg:val}
\end{algorithm}
\end{minipage}
~
\begin{minipage}[t]{0.39\textwidth}
\begin{algorithm}[H]
\small

  \caption{\textsf{rob}$(\tau, \varphi)$}
\begin{algorithmic}[1]

\State $\waut_{\varphi} \leftarrow \textsf{T}(\varphi)$
\State $\waut_{\neg \varphi} \leftarrow \textsf{T}(\neg \varphi)$
\State $v_{1} \leftarrow \textsf{val}(\tau, \waut_{\varphi})$
\State $v_{2} \leftarrow \textsf{val}(\tau, \waut_{\neg \varphi})$

\If{$v_{1} = \oone$  }
	\Return $v_2$
\Else
	~\Return $-v_1$
\EndIf
\end{algorithmic}
\label{alg:rob}
\end{algorithm}
\end{minipage}

We now state that Algorithm~\ref{alg:val} correctly computes the value of $\trace$ in $\waut$, which corresponds to the 
distance $d(\trace, \spec)$. 
\begin{theorem}
\label{th:val}
Given a specification $\varphi$, its associated SWA $\waut$ defined over a semiring $S$ and a trace $\trace$, we have that 
$\textsf{val}(\trace, \waut) = \tvalue(\trace, \waut) = d(\trace, \varphi)$.
\end{theorem}



We finally build the monitor that measures the robustness degree between the trace $\trace$ and a specification $\varphi$ 
by computing the value of $\tau$ in $\waut_{\varphi}$ and 
$\waut_{\neg \varphi}$.  This procedure is summarized in Algorithm~\ref{alg:rob} that trivially implements the robustness measure 
$\rho$. We show that our abstract computation of $\rho$ is sound and complete.

\vspace{-4pt}
\begin{theorem}[Soundness and completeness]
\label{th:qual}
Given traces $\trace$ and $\trace'$, a specification $\varphi$ and distances $d(\tau, \tau')$, $d(\tau, \neg \varphi)$ defined over a bounded semiring $S$, 
\vspace{-10pt}
$$
\begin{array}{rcl}
\rho(\trace, \varphi) > 0 & \rightarrow & \tau \models \varphi \\
\rho(\trace, \varphi) < 0 & \rightarrow & \tau \not \models \varphi \\
\trace \models \spec \textrm{ and } d(\trace, \trace') \sqsubset d(\trace, \neg \spec) & \rightarrow & \trace' \models \spec.
\end{array}
$$
\end{theorem} 
\vspace{-4pt}

We first observe that if $\rho(\trace, \varphi) = 0$, then we do not know (only from that number) 
whether $\trace$ satisfies $\spec$. We illustrate this observation with the formula $x > 0$ and 
the trace $\tau = 0$ of size one. It is clear that $\tau \not \models \spec$ but the actual distance between the element in 
$\{ v~|~v > 0 \}$ that is closest to $0$ and $0$ is infinitesimally close to $0$. 
In order to have both directions of the implications in the soundness proof and 
to guarantee that the robustness degree is never equal to $0$, we would need to introduce non-standard reals. Note that even with the 
current setting, we can easily say whether $\trace \models \spec$, even when $\rho(\tau, \spec) = 0$. Second, 
we do not need to explicitly compute the complement automaton $\waut_{\neg \varphi}$ if $\waut_{\varphi}$ is deterministic. In that case, 
it is sufficient to apply a slight modification of Algorithm~\ref{alg:val} on $\waut_{\varphi}$ only. The modification consists in 
reporting either the minimum value of an accepting or the minimum value of a non-accepting location, depending on whether 
the trace satisfies $\spec$.

\vspace{-12pt}
\subsubsection{Complexity of the translation and the algorithm}

The number of locations in the symbolic automaton $\waut_{\varphi}$ is exponential in the size of the formula 
$\varphi$ for both STL and SRE (we recall that we use SRE with the intersection operator). The size of the predicates decorating the 
transitions in $\waut_{\varphi}$ is also exponential in the number of propositions appearing in the formula, due to the translation of the 
predicate to DNF form. The algorithm $\textsf{val}(\trace, \waut)$ applied to a trace $\tau$ of size $l$, and $\waut$ with $n$ locations, 
$m$ transitions, and maximum size $p$ of the predicates, takes in the order of $l\,{\cdot}\,p\,{\cdot}\,(\max(m,n))$ iterations. 
 
\vspace{-14pt}

\subsection{Instantiating Monitors}
\label{sec:inst}

\vspace{-4pt}
In Sections~\ref{sec:distance} and~\ref{sec:trval}, we presented an abstract monitoring procedure that measures a robustness degree of a trace $\tau$ 
with respect to a specification $\varphi$. We give concrete semantics to these monitors by instantiating the semiring and the distance function. Here we consider 
three instantiations of the procedure:
\vspace{-4pt}
\begin{enumerate}
\item \label{one} Boolean semiring with $d(a,b) = 1$ if $a \neq b$ and $0$ otherwise
\item \label{two} Minmax semiring with $d(a,b) = |a - b|$
\item \label{three} Tropical semiring with $d(a,b) = |a - b|$
\end{enumerate}
\vspace{-4pt}
The instantiation~\ref{one} gives the monitors classical qualitative semantics, where the distance of $\trace$ from $\varphi$ is $0$ if $\trace$ is in the language of $\varphi$ and 
$1$ otherwise. The instantiation~\ref{two} gives the computation of the robustness degree based on the infinite norm, as defined in~\cite{fainekos-robust}. The 
instantiation~\ref{three} gives the computation of the robustness degree based on the Hamming distance lifted to the sets.

\vspace{-4pt}
\begin{example}
We illustrate our monitoring procedure instantiated to different semantics in Table~\ref{tab:exampleWSA}. We choose the trace 
$\trace = (4,2) \cdot (5,3) \cdot (2,5) \cdot (3,5)$ that violates the specifications $\varphi_{1}$ and $\varphi_{2}$ from Example~\ref{ex:spec}. We instantiate 
Algorithm~\ref{alg:val} to the specific semirings and apply each instantiation to the $\trace$ and $\waut$ from Figure~\ref{fig:exampleWSA}. We mark in bold the accepting state 
and the trace value.
\end{example}

\vspace{-12pt}
\begingroup
\begin{table}

\centering
\begin{tabular}{||c|c|c|c|c|c|c||}
\cline{1-7}
 semiring &  ~state~ &  ~init~  &  ~$v_0 = (4,2)$~  & ~$v_1 = (5,3)$~  & ~$v_2 = (2,5)$~  &  ~$v_3 = (3,5)$~\\
\cline{1-7}
\cline{1-7}
\multirow{3}{*}{~Boolean~}  
	& $q_0$ 		  & 	  ~0~&~0~&~0~& ~0~&~0~ \\
\cline{2-7}
	& $q_1$ 		  &    ~1~&~1~&~1~&~1~&~1~ \\
\cline{2-7}
	& \boldmath$q_2$ &    ~1~& ~1~&~1~&~1~&~\textbf{1}~ \\

\clineB{1-7}{2.5}
\multirow{3}{*}{~MinMax~}  
	& $q_0$ 		  & 	  ~0~&~0~&~0~&~0~&~0~ \\
\cline{2-7}
	& $q_1$ 		  &~$\infty$~&~4~&~3~&~1~&~1~ \\
\cline{2-7}
	& \boldmath$q_2$ &~$\infty$~&~$\infty$~&~4~&~3~&~\textbf{1}~ \\

\clineB{1-7}{2.5}
\multirow{3}{*}{~Tropical~}  
	& $q_0$ 		  &~0~&~0~&~0~&~0~&~0~ \\
\cline{2-7}
	& $q_1$ 		  &~$\infty$~&~5~&~5~&~2~&~1~ \\
\cline{2-7}
	& \boldmath$q_2$ &~$\infty$~&~$\infty$~&~10~&~7~& ~\textbf{3}~ \\
\cline{1-7}
\end{tabular}
\vspace{2ex}

\setlength{\belowcaptionskip}{-8pt}
\caption{\textsf{val}$(\tau, \waut)$ computed on SWA from Figure~\ref{fig:exampleWSA} with different semirings.}
\label{tab:exampleWSA}
\normalsize
\end{table}
\endgroup

\vspace{8pt}

We note that by Theorem~\ref{th:val}, our computation of robustness is {\em precise} with respect to the semantics of the specification, regardless of the instantiated semiring. This is in contrast to the syntactic approaches to robustness~\cite{fainekos-robust,robust1} that under-approximate the 
robustness value. In case we instantiate MinMax seminiring in ARV, we do not demonstrate imprecision shown in Examples 17,18 from~\cite{fainekos-robust}. The comparison results from section~\ref{eval:avcs} confirm this observation.

%% file: steps.pdf_t
\begin{picture}(0,0)%
\includegraphics{steps.pdf}%
\end{picture}%
\setlength{\unitlength}{3947sp}%
\begingroup\makeatletter\ifx\SetFigFont\undefined%
\gdef\SetFigFont#1#2#3#4#5{%
  \reset@font\fontsize{#1}{#2pt}%
  \fontfamily{#3}\fontseries{#4}\fontshape{#5}%
  \selectfont}%
\fi\endgroup%
\begin{picture}(7524,1824)(1489,-3373)
\put(2401,-1861){\makebox(0,0)[b]{\smash{{\SetFigFont{9}{10.8}{\rmdefault}{\mddefault}{\updefault}{\color[rgb]{0,0,0}Regular Specification}%
}}}}
\put(8101,-1711){\makebox(0,0)[b]{\smash{{\SetFigFont{9}{10.8}{\rmdefault}{\mddefault}{\updefault}{\color[rgb]{0,0,0}Symbolic Weighted }%
}}}}
\put(8101,-1861){\makebox(0,0)[b]{\smash{{\SetFigFont{9}{10.8}{\rmdefault}{\mddefault}{\updefault}{\color[rgb]{0,0,0}Automaton}%
}}}}
\put(8101,-2086){\makebox(0,0)[b]{\smash{{\SetFigFont{12}{14.4}{\rmdefault}{\mddefault}{\updefault}{\color[rgb]{0,0,0}$\weighted_{\varphi}$}%
}}}}
\put(5251,-1711){\makebox(0,0)[b]{\smash{{\SetFigFont{9}{10.8}{\rmdefault}{\mddefault}{\updefault}{\color[rgb]{0,0,0}Symbolic}%
}}}}
\put(5251,-1861){\makebox(0,0)[b]{\smash{{\SetFigFont{9}{10.8}{\rmdefault}{\mddefault}{\updefault}{\color[rgb]{0,0,0}Automaton}%
}}}}
\put(5251,-2086){\makebox(0,0)[b]{\smash{{\SetFigFont{12}{14.4}{\rmdefault}{\mddefault}{\updefault}{\color[rgb]{0,0,0}$\aut_{\varphi}$}%
}}}}
\put(2401,-2086){\makebox(0,0)[b]{\smash{{\SetFigFont{12}{14.4}{\rmdefault}{\mddefault}{\updefault}{\color[rgb]{0,0,0}$\varphi$}%
}}}}
\put(8101,-3061){\makebox(0,0)[b]{\smash{{\SetFigFont{9}{10.8}{\rmdefault}{\mddefault}{\updefault}{\color[rgb]{0,0,0}Distance}%
}}}}
\put(8101,-3286){\makebox(0,0)[b]{\smash{{\SetFigFont{12}{14.4}{\rmdefault}{\mddefault}{\updefault}{\color[rgb]{0,0,0}$d(\trace,\varphi)$}%
}}}}
\put(2401,-3061){\makebox(0,0)[b]{\smash{{\SetFigFont{9}{10.8}{\rmdefault}{\mddefault}{\updefault}{\color[rgb]{0,0,0}Trace}%
}}}}
\put(2401,-3286){\makebox(0,0)[b]{\smash{{\SetFigFont{12}{14.4}{\rmdefault}{\mddefault}{\updefault}{\color[rgb]{0,0,0}$\trace$}%
}}}}
\end{picture}%

%% file: example.tex
$$ 
\begin{tikzpicture}[->,>=stealth',initial text=,shorten >=1pt,auto,node distance=4.5cm,
                    semithick]
  \tikzstyle{every state}=[fill=none,draw=black,text=black]

  \node[initial ,state]	      (0)  			     {$q_0$};
  \node[state]    (1)  [right of=0] 			{$q_1$}; 
  \node[state, accepting]    (2)  [right of=1] {$q_2$};
  \node[state, draw=white] () [below = 5.0em] at (2.west) {};

  \path 
  	  (0) edge [loop above]  node {$\top$} (0)
    	  (0) edge    node [align=center] {$ x \leq 5 \wedge x \leq 3 \wedge y \geq  6$} (1)
    	  (0) edge    node [align=center, below] {$\delta_1$} (1)
    	  (0) edge    node [draw, align=center, yshift=0.75cm] {$ x \leq 3 \wedge y \geq  6$} (1)
   	  (1) edge  node [align=center] {$x \leq 3 \wedge y \geq  6$} (2)
   	  (1) edge  node [draw, align=center, yshift=0.75cm] {$x \leq 3 \wedge y \geq  6$} (2)
    	  (1) edge    node [align=center, below] {$\delta_2$} (2)
  	  (2) edge [loop above] node {$\top$} (2);
\end{tikzpicture}
$$ 

%% file: idem.pdf_t
\begin{picture}(0,0)%
\includegraphics{idem.pdf}%
\end{picture}%
\setlength{\unitlength}{3947sp}%
\begingroup\makeatletter\ifx\SetFigFont\undefined%
\gdef\SetFigFont#1#2#3#4#5{%
  \reset@font\fontsize{#1}{#2pt}%
  \fontfamily{#3}\fontseries{#4}\fontshape{#5}%
  \selectfont}%
\fi\endgroup%
\begin{picture}(3930,1567)(2086,-4616)
\put(3301,-4561){\makebox(0,0)[b]{\smash{{\SetFigFont{12}{14.4}{\rmdefault}{\mddefault}{\updefault}{\color[rgb]{0,0,0}$2$}%
}}}}
\put(3901,-4561){\makebox(0,0)[b]{\smash{{\SetFigFont{12}{14.4}{\rmdefault}{\mddefault}{\updefault}{\color[rgb]{0,0,0}$3$}%
}}}}
\put(4501,-4561){\makebox(0,0)[b]{\smash{{\SetFigFont{12}{14.4}{\rmdefault}{\mddefault}{\updefault}{\color[rgb]{0,0,0}$4$}%
}}}}
\put(5101,-4561){\makebox(0,0)[b]{\smash{{\SetFigFont{12}{14.4}{\rmdefault}{\mddefault}{\updefault}{\color[rgb]{0,0,0}$5$}%
}}}}
\put(5701,-4561){\makebox(0,0)[b]{\smash{{\SetFigFont{12}{14.4}{\rmdefault}{\mddefault}{\updefault}{\color[rgb]{0,0,0}$6$}%
}}}}
\put(5401,-4036){\makebox(0,0)[b]{\smash{{\SetFigFont{12}{14.4}{\rmdefault}{\mddefault}{\updefault}{\color[rgb]{0,0,0}$1$}%
}}}}
\put(4801,-3736){\makebox(0,0)[b]{\smash{{\SetFigFont{12}{14.4}{\rmdefault}{\mddefault}{\updefault}{\color[rgb]{0,0,0}$3$}%
}}}}
\put(4801,-3436){\makebox(0,0)[b]{\smash{{\SetFigFont{12}{14.4}{\rmdefault}{\mddefault}{\updefault}{\color[rgb]{0,0,0}$3$}%
}}}}
\put(6001,-4561){\makebox(0,0)[b]{\smash{{\SetFigFont{12}{14.4}{\rmdefault}{\mddefault}{\updefault}{\color[rgb]{0,0,0}$x$}%
}}}}
\put(5701,-3286){\makebox(0,0)[b]{\smash{{\SetFigFont{12}{14.4}{\rmdefault}{\mddefault}{\updefault}{\color[rgb]{0,0,0}$v(x)$}%
}}}}
\put(2101,-4561){\makebox(0,0)[b]{\smash{{\SetFigFont{12}{14.4}{\rmdefault}{\mddefault}{\updefault}{\color[rgb]{0,0,0}$0$}%
}}}}
\put(2701,-4561){\makebox(0,0)[b]{\smash{{\SetFigFont{12}{14.4}{\rmdefault}{\mddefault}{\updefault}{\color[rgb]{0,0,0}$1$}%
}}}}
\put(3001,-3886){\makebox(0,0)[b]{\smash{{\SetFigFont{12}{14.4}{\rmdefault}{\mddefault}{\updefault}{\color[rgb]{0,0,0}$x \leq 3$}%
}}}}
\put(3601,-4186){\makebox(0,0)[b]{\smash{{\SetFigFont{12}{14.4}{\rmdefault}{\mddefault}{\updefault}{\color[rgb]{0,0,0}$x \leq 5$}%
}}}}
\put(3001,-3586){\makebox(0,0)[b]{\smash{{\SetFigFont{12}{14.4}{\rmdefault}{\mddefault}{\updefault}{\color[rgb]{0,0,0}$x \leq 3 \wedge x \leq 5$}%
}}}}
\end{picture}%

%% file: eval.tex
\section{Case Studies}
\label{sec:evaluation}

\vspace{-8pt}

We implemented our approach (ARV) in a prototype tool in Java. In order to determine satisfiability of SWA transition constraints, we used the Z3 solver~\cite{de2008z3}. We evaluate our framework on two case studies from the automotive domain: The Autonomous Vehicle Control Stack model~\cite{alena_bench} and the Automatic Transmission System model~\cite{hoxha_bench}. In the first case study we also compare the precision of ARV with relevant tools developed by the RV community. 

\vspace{-8pt}
\subsection{Autonomous Vehicle Control Stack}
\label{eval:avcs}

\vspace{-4pt}
The first benchmark is a model of an autonomous vehicle control stack, which is used to solve the trajectory planning problem. The stack consists of three layers, starting from the top: Behavioral Planner (BP), Trajectory Planner (TP) and Trajectory Tracker (TT). The BP provides the coarse-grain trajectory way-points for the Autonomous Vehicle (AV), and supplies them to the underlying  TP which calculates fine grain trajectory points, using cubic spline trajectory generation. The lowest layer is the TT which actuates the AV based on the trajectory points in order to steer it towards the requested path. 

\begin{figure}[h!]
\centering
\includegraphics[width=1\textwidth]{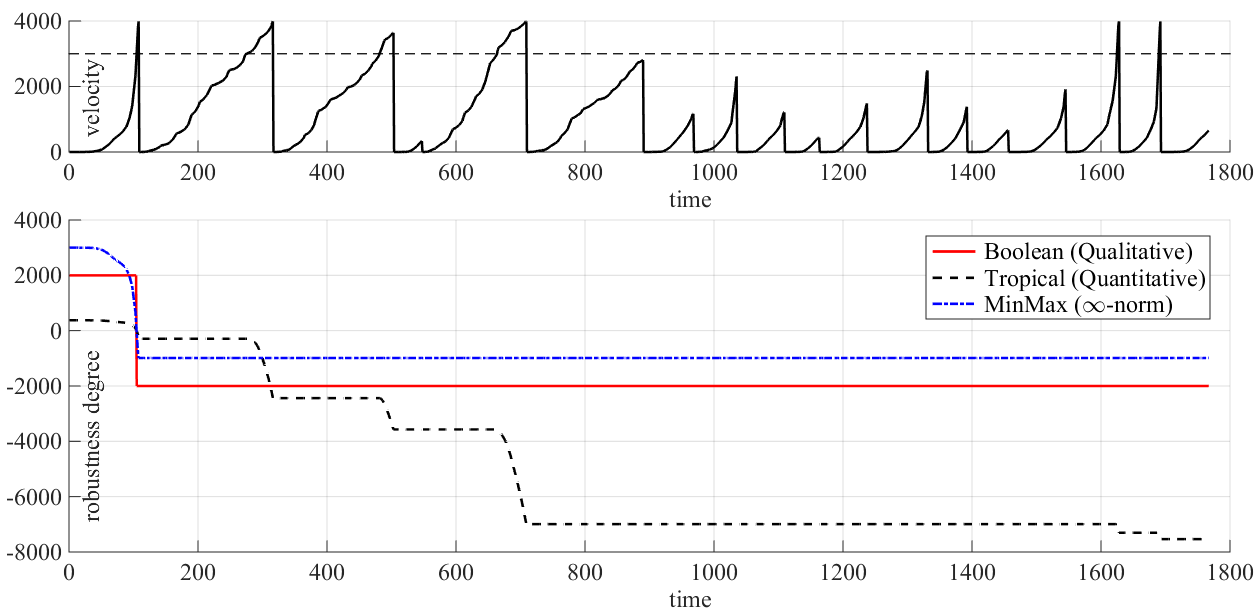}
\setlength{\abovecaptionskip}{-2pt}
\setlength{\belowcaptionskip}{-6pt}
\caption{Robustness degree $\robustness(\trace_{[0,t]}, \varphi_{1} )$, where $\varphi_{1} = \square(v_{ego} \leq v_{limit})$, based on three different semiring instantiations.}
\label{fig:f2}
\end{figure}

\begin{figure}[hb!]
\centering
\includegraphics[width=1\textwidth]{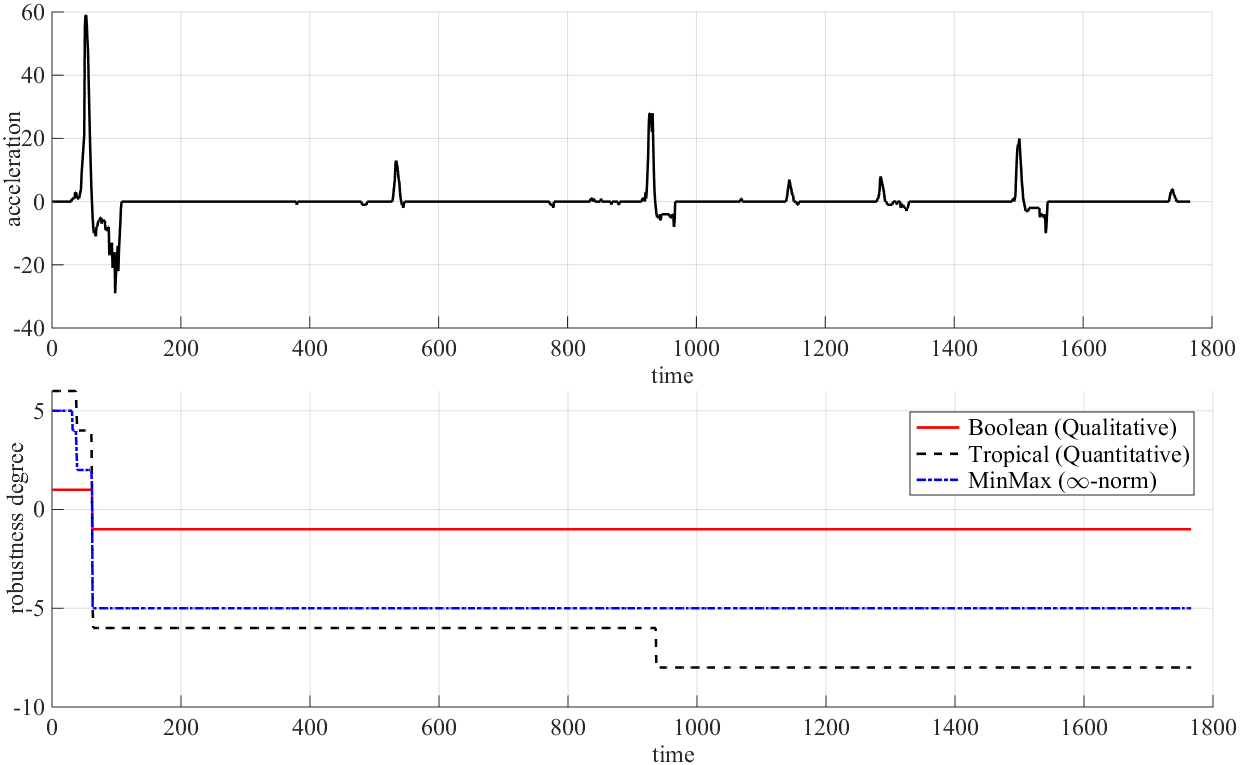}
\setlength{\abovecaptionskip}{-2pt}
\setlength{\belowcaptionskip}{-6pt}
\caption{Robustness degree $\robustness(\trace_{[0,t]}, \varphi_{2} )$, where $ \varphi_{2} = \square((a_{x} \geq \theta) \rightarrow \square_{(0,\epsilon]} \neg (a_{x} \leq 0) )$, based on different semiring instantiations.}
\label{fig:f3}
\end{figure}

The benchmark supports three distinct layouts: a room with obstacles, a curved road and a roundabout. The obstacles and undesired areas are determined by assigning the specific cost. The autonomous vehicles can simulate 4 scenarios:
parallel driving, paths crossing without collision and collision avoidance by either the first or the second vehicle. We ran the scenario of AVs performing parallel driving on roundabout layout and we obtained the traces for the speed $v_{x}$ and acceleration $a_{x}$. We specified two requirements, which model normal operation (w.r.t. acceleration) and traffic rules (speed limit).

In Fig~\ref{fig:f2} we demonstrate various robustness degrees for the following requirement:
``The ego vehicle travels at a velocity less than or equal to the speed limit''. We formalize this requirement in STL:
$ \varphi_{1} = \G(v_{ego} \leq v_{limit})$. In Fig~\ref{fig:f3} we monitor the following requirement
``If the autonomous vehicle started to accelerate, then it will not start decelerating in the very near future''. We formalize this requirement in STL:
$ \varphi_{2} = \G((a_{x} \geq \theta) \rightarrow \G_{(0,\epsilon]} \neg (a_{x} \leq 0) )$. We select $ \theta = 5$ and $ \epsilon = 3$.
For the clarity of the graph, some values are (down)scaled.

\vspace{-12pt}
\subsubsection{Comparison with S-TaLiRo and Breach}
Both S-TaLiRo and Breach implement robustness monitoring algorithms, in which robustness is measured 
with infinite norm. The algorithms implemented in these tools are syntactic, i.e. they work directly (and inductively) on the structure of the formula without 
passing via automata. In this section, we intend to demonstrate the preciseness of our semantic automata-based approach. We note that in contrast to our 
setting, these two tools work with continuous time. In order to enable the comparison, we instantiate our monitors to the MinMax semiring and 
we simulate the Autonomous Vehicle Control Stack model with fixed sampling, 
thus ensuring that the discrete versus continuous discrepancy does not affect the results.



In $\psi_{1}$ and $\psi_{2}$ we test the sensitivity of the robustness degree algorithm w.r.t. to the minimal set representation. 
The formula $\psi_{1}$ defines the acceptable range $[-30,30]$ of values for $a$, while $\psi_{2}$ is a semantically equivalent formula that represents the 
same set as a disjoint union $[-30,0) \cup [0, 30]$. Similarly, $\psi_{3}$ and $\psi_{4}$ represent two semantically equivalent temporal formulas. Finally, 
both $\psi_{5}$ and $\psi_{6}$ represent formulas that are unsatisfiable. We can observe that our approach produces results that are invariant to the syntactic 
representation of the formula. In particular, our monitoring algorithm is able to detect unsatisfiable formulas. In contrast, neither S-TaLiRo nor Breach can 
detect unsatisfiable specifications. We can also observe that in some cases, these two tools are also sensitive to the syntactic representation of the specification. 
This is visible in the computation of the robustness for $\psi_{1}$ and $\psi_{2}$, where S-TaLiRo computes the inconclusive result $0$ for a specification that is 
satisfied by the trace, while Breach correctly finds that the formula is satisfied, but assigns it a very low robustness degree.



\vspace{-12pt}

\begingroup
\renewcommand*{\arraystretch}{1.2}
\begin{table}

\centering
\begin{tabular}{||lc|c|c|c||}
\cline{1-5}
 			&  & ~S-TaLiRo~  & ~Breach~ & ~ARV~ \\

\cline{1-5}
~$\psi_{1}=~$ & ~$a \geq -30 \wedge a \leq 30$~                                                & 30   &  30                           & 30                    \\
~$\psi_{2}=~$ & ~$(a \geq -30 \wedge a < 0) \vee (a \geq 0 \wedge a \leq 30) $ ~ &  0    &     $10^{-13} $         & 30                    \\
\hline
~$\psi_{3}=~$ & ~$\F(a \geq -10)$~                                                                          &69    & 69      & 69        \\
~$\psi_{4}=~$ & ~$\F((a \geq -10 \wedge a \leq 60) \vee (a \geq 55))$ ~                     &35   &  35       & 69        \\
\hline
~$\psi_{5}=$~ & ~$\G(a \geq 5  \wedge  a < 5) $ ~                                            &-64    & -59        & $-\infty$                    \\
~$\psi_{6}=$~ & ~$\neg(\F \psi_{1} \vee   \F(a < -30  \vee  a > 30)) $ ~             &-30    & -30        & $-\infty$                   \\
\cline{1-5}

\end{tabular}
\vspace{2ex}

\caption{Precision comparison between syntactic and semantic approach.}
\label{tab:precise}
\normalsize
\end{table}
\endgroup



\vspace{-36pt}
\subsection{Automatic Transmission System}

We first consider the Automatic Transmission deterministic model~\cite{hoxha_bench} as 
our system-under-test (SUT). It is a model of a transmission controller that exhibits both continuous 
and discrete behavior. The system has two inputs -- the throttle $u_{t}$ and the break $u_{b}$. The break allows the user to 
model variable load on the engine. The system has two continuous-time state variables -- the speed of the 
engine $\omega$ (RPM), the speed of the vehicle $v$ (mph) and the active gear $g_{i}$. 


The system is initialized with zero 
vehicle and engine speed. It follows that the output trajectories depend only on the input signals $u_{t}$ and $u_{b}$, which can 
take any value between $0$ and $100$ at any point in time. The Simulink model contains $69$ blocks including $2$ integrators, 
$3$ look-up tables, $2$ two-dimensional look-up tables and a Stateflow chart with $2$ concurrently executing finite state machines 
with $4$ and $3$ states, respectively. The benchmark~\cite{hoxha_bench} defines $8$ STL formalized requirements that the system must satisfy. 

We select the following requirement for our case study: ``The engine and the vehicle speed never reach $\omega_{max}$ and $v_{max}$''.
We use STL to formalize this requirement as follows:~$\varphi_{3} = \G((\omega \leq \omega_{max})~\land~(v \leq v_{max})).$

\begin{figure}[h!]
\centering
\includegraphics[width=1\textwidth]{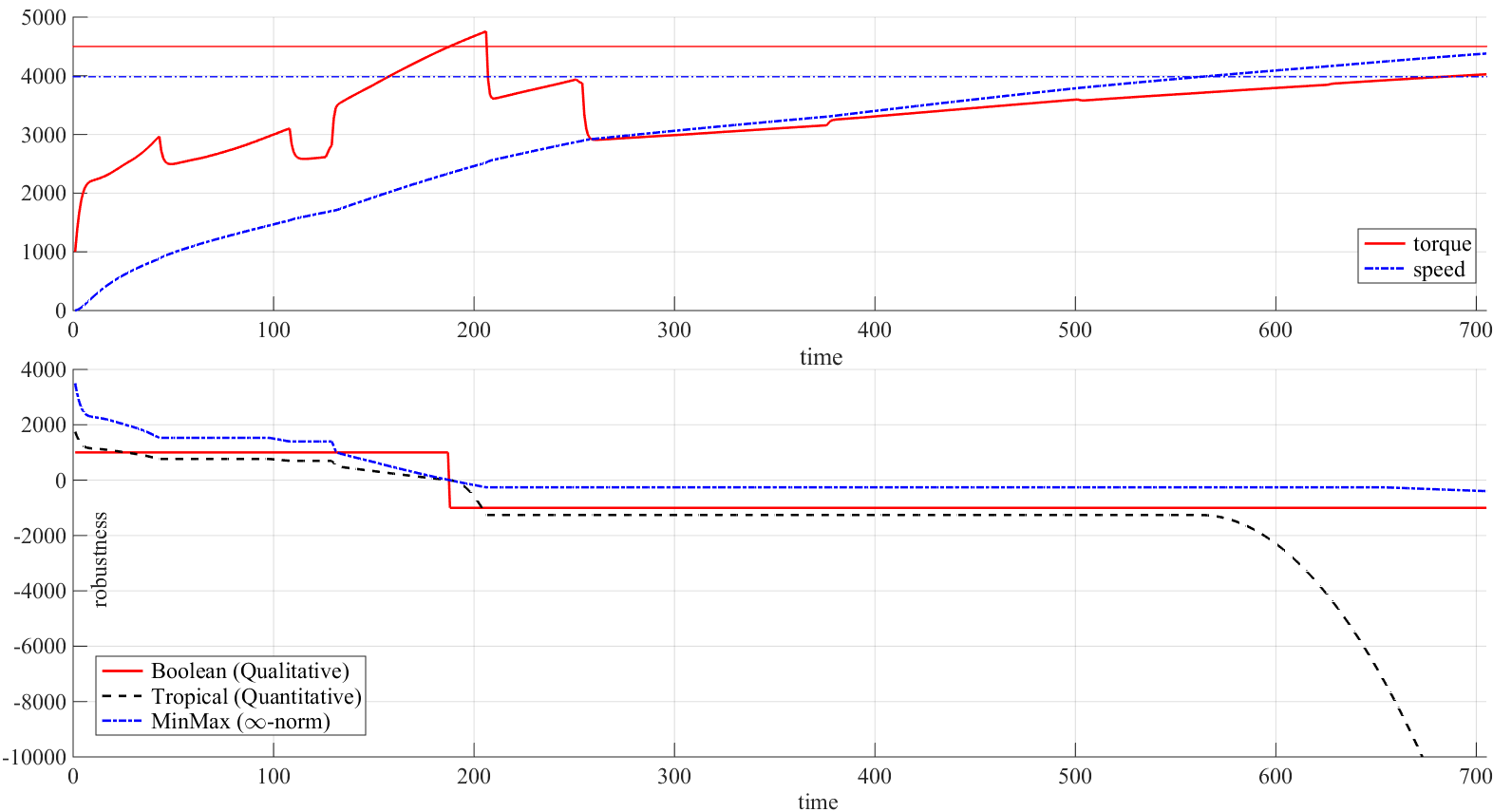}
\caption{Robustness degree $\robustness(\trace_{[0,t]}, \varphi_{3} )$, where $ \varphi_{3} =  \square((\omega \leq \omega_{max})~\land~(v \leq v_{max}))$, based on different semiring instantiations.}

\label{fig:f1}
\end{figure}

We translate $\varphi_{3}$ into a SWA and instantiate it with three semirings -- Boolean, MinMax and Tropical, thus obtaining monitors for qualitative, $\infty$-norm and Hamming distance based quantitative semantics. We note that the qualitative semantics yields the binary verdicts. In contrast, the quantitative semantics based on the $\infty$-norm is obtained by instantiating MinMax idempotent semiring. Therefore the robustness degree based of this semantics relies on maximum pointwise distance, without accumulating it over time. For this reason we find it useful for performing worst-case analysis. Finally, the quantitative semantics based on Tropical semiring accumulates the pointwise distances over the entire trace due to non-idempotent $\otimes$ (addition). Thus, the robustness degree in each time instance depends on the robustness of the entire trace prefix, ensuring that no information on robustness is lost over time. We can finally observe that all the quantitative semantics are consistent with qualitative semantics, as stated in Theorem~\ref{th:qual}. This is observable in Fig~\ref{fig:f1}, before time reaches 200.
\vspace{-8pt}


%% file: conclusion.tex
\section{Conclusions and Future Work}
\label{sec:con}

We presented ARV, our generic algebraic approach to monitoring correctness and robustness of CPS applications. 
We demonstrated the flexibility of ARV with respect to specification languages and semantics. We showed that 
ARV, which relies on the use of automata, enables a precise robustness measurement of a trace with respect to 
a specification. We also believe that defining the abstract RV algorithm over the automaton structure will facilitate 
code generation of real-time monitors for various platforms, including software and FPGAs.

We believe that the results presented in this paper may open many new research and development avenues. In this paper, 
we have an enumerative approach to real-time. We will investigate the effect of symbolic representation of time to our automata-based 
approach. 

We have seen that the precision of our semantic approach comes at a price -- an exponential blow up in the number of locations and 
the size of the transition predicates. In this paper, we studied the worst-case complexity 
of our translation. However, we believe that for many applications the effective translation will yield monitors that are 
much smaller from the worst case. This requires a comprehensive experimental study with an optimized implementation. 
To achieve this goal, we will explore different optimization strategies, including the 
use of the algorithms implemented in the symbolic automata 
library\footnote{\url{https://github.com/lorisdanto/symbolicautomata}} such as its minimization  
procedure~\cite{simmin}. We will integrate our robustness monitoring approach to the existing falsification testing 
frameworks and quantify the improvements due to the preciseness of our algorithms. We will implement several code generators 
(interpreted Java, Simulink S-functions, embedded C, FPGA, etc.) and investigate the reuse of specifications and monitors 
across stages in the development cycle.

In this paper, we restricted ourselves to Hausdorff-like measures, and additive idempotent semirings. We would like to
study in the future, the extension of our framework to non-idempotent $\oplus$-addition, and its application to RV: For
instance, to a probabilistic semiring. We would also like to investigate whether we can use our approach to support other types
of semantics.  For instance, we will study whether we can generalize ARV to enable measuring weighted edit distance measure between 
a trace and a specification as proposed in~\cite{stledit}.

%% file: specs.tex
\section{Specification Languages}
\label{sec:spec}

In this appendix, we recall the syntax and semantics of STL and SRE, both 
interpreted over discrete time.

\subsection{Signal Temporal Logic}

We consider $\stl$ with both {\em past} and {\em future} operators interpreted over 
digital signals of final length.
We assume that $\domain$ is a metric space equiped with a distance $d$.
The syntax of a $\stl$ formula $\varphi$ over $X$ is defined by the grammar:
$$
\begin{array}{lcl}
\varphi & := & x \sim u~|~\neg \varphi~|~\varphi_1 \vee \varphi_2~|~\varphi_1 \until_I \varphi_2~|~\varphi_1 \since_I \varphi_2 \\
\end{array}
$$
\noindent where $x \in X$, $\sim \in \{ <, \leq \}$, $u \in \mathbb{D}$, 
$I$ is of the form $[a,b]$ or $[a, \infty)$ such that $a,b \in \mathbb{N}$ and 
$0 \leq a \leq b$. The other standard operators are derived as follows:
$\true = p \vee \neg p$, $\false = \neg \true$, $\varphi_1 \wedge \varphi_2 = \neg(\neg \varphi_1 \vee 
\neg \varphi_2)$, $\F_I \varphi = \true \until_I \varphi$, $\G_I \varphi = \neg \F_I \neg \varphi$, 
$\opP_I \varphi = \true \since_I \varphi$, $\opH_I  \varphi = \neg \opP_I \neg \varphi$, 
$\X \varphi = \false \until_{[1,1]} \varphi$ and  $\Y \varphi = \false \since_{[1,1]} \varphi$.

The semantics of a $\stl$ formula with respect to a signal $s$ of length $l$ is described via the 
satisfiability relation $(s,i) \models \varphi$, indicating that the signal $s$ satisfies $\varphi$ at the 
time index $i$, according to the following definition where $\mathbb{T} = [0, l)$.
$$
\begin{array}{lcl}
(s,i) \models x \sim u & \leftrightarrow & s(i,x) \sim u \\
(s,i) \models \neg \varphi & \leftrightarrow & (s,i) \not \models \varphi \\
(s,i) \models \varphi_1 \vee \varphi_2 & \leftrightarrow & (s,i) \models \varphi_1 \; \textrm{or} \; (s,i) \models \varphi_2 \\
(s,i) \models \varphi_1 \until_I \varphi_2 & \leftrightarrow & \exists j \in (i + I) \cap \mathbb{T} \; \textrm{:} \;
(s,j) \models \varphi_2 \; \textrm{and} \; \forall i < k < j, (s,k) \models \varphi_1 \\
(s,i) \models \varphi_1 \since_I \varphi_2 & \leftrightarrow & \exists j \in (i - I) \cap \mathbb{T} \; \textrm{:} \;
(s,j) \models \varphi_2 \; \textrm{and} \; \forall j < k < i, (s,k) \models \varphi_1 \\
\end{array}
$$

We note that we use the semantics for  $\since_I$ and $\until_I$ that is strict in both arguments and that we allow 
punctual modalities due to the discrete time semantics.
Given an $\stl$ formula $\varphi$, we denote by $L(\varphi)$ the {\em language} of $\varphi$, which is the 
set of all signals $s$ such that $(s,0) \models \varphi$.

\subsection{Signal Regular Expressions}
\label{subsec:SRE}

Signal Regular Expressions (SRE)~\cite{OdedSRE} allow to pattern-match 
a specification over a signal. As the authors in~\cite{measures}
mentioned, the fundamental difference between STL and SREs comes from a
fact that the satisfaction of an STL formula is computed for a time
point, while the match of a SRE results in a time interval.
In this work we adapt the definition of SREs from~\cite{measures}
with an assumption of interpreting SREs over discrete time.
The syntax of a SRE formula $\varphi$ over $X$ is defined by the grammar:
\vspace{-4pt}
 $$
\begin{array}{lcl}
b & := & x \sim u~|~\neg b~|~b_{1} \vee b_{2} \\
\varphi & := & \epsilon~|~b~|~\varphi_{1} \cdot \varphi_{2}~|~\varphi_{1} \cup \varphi_{2}~|~\varphi_1 \cap \varphi_2~|~\varphi^{*}~|~\langle \varphi \rangle_{I} \\
\end{array}
$$

\noindent where $x \in X$, $\sim \in \{ <, \leq \}$, $u \in \mathbb{D}$, 
$I$ is of the form $[a,b]$ or $[a, \infty)$ such that $a,b \in \mathbb{N}$ and 
$0 \leq a \leq b$. Although we interpret SRE over discrete time, we interpret its operators following the style of continuous time TRE. As a consequence, 
a signal segment that matches a predicate such as $x \leq 5$ means that it matches it for a strictly positive duration. The time duration operator $\langle \varphi \rangle_{I}$ 
is matched by a segment if it has a duration in $I$.

The semantics of SRE $\varphi$ with respect to discrete signal $w$ and time instances $i \leq i'$ is given
in terms of satisfaction relation $(w,i,i') \models \varphi$:
$$
\begin{array}{lcl}
(w,i,i') \models \epsilon & \leftrightarrow & i = i' \\
(w,i,i') \models q          & \leftrightarrow & i \leq i' \; \textrm{and} \; \forall i'' \textrm{ s.t. }  i \leq i'' < i', \pi_p(w)[i''] = 1  \\
(w,i,i') \models \varphi_1 \cdot \varphi_2 & \leftrightarrow & \exists i'' \textrm{ s.t. } i \leq i'' < i', (w,i,i'') \models \varphi_1 \; \textrm{and} \; (w, i'', i') \models \varphi_2 \\
(w,i,i') \models \varphi_1 \cup \varphi_2 & \leftrightarrow & (w,i,i') \models \varphi_1 \; \textrm{or} \; (w, i, i') \models \varphi_2 \\
(w,i,i') \models \varphi_1 \cap \varphi_2 & \leftrightarrow & (w,i,i') \models \varphi_1 \; \textrm{and} \; (w, i, i') \models \varphi_2 \\
(w,i,i') \models \varphi^{*}                   & \leftrightarrow & (w,i,i') \models \epsilon \; \textrm{or} \; (w,i,i') \models \varphi \cdot \varphi^{*} \\
(w,i,i') \models \langle \varphi \rangle_I  & \leftrightarrow & i' - i \in I \; \textrm{and} \; (w, i, i') \models \; \varphi \\
\end{array}
$$

%% file: proof_th1.tex
\newpage

\section{Proofs}
\label{sec:correctness}

\begin{customlem}{\ref{lemma:one-min}}
Let $\algebra = (S, \oplus, \otimes, \ozero, \oone)$ be an additively idempotent, negative and monotonic semiring. Then, for all 
$a \in S$, $\oone \sqsubseteq a \sqsubseteq \ozero$.
\end{customlem}

\begin{proof}
Consider an arbitrary $a \in S$. By Definition~\ref{def:mono} and 
assumption that $S$ is negative, we have that $\oone \sqsubseteq \ozero$. By Definition~\ref{def:mono} and assumption that 
$S$ is monotonic, we have that $\oone \otimes a \sqsubseteq \ozero \otimes a$ and $\oone \oplus a \sqsubseteq \ozero \oplus a$. 
By Definition~\ref{def:semiring}, $\oone \sqsubseteq a$ and $a \sqsubseteq \ozero$, which concludes the proof.
\end{proof}

\begin{customthm}{\ref{thm:valpred}}

Given a predicate $\predicate$ in DNF, a valuation $v$ and the distance 
$d(v, \predicate)$ defined over a bounded semiring $S$, we have that $\textsf{vpd}(v,\predicate) = d(v, \predicate)$ 
if: (1) $S$ is idempotent; or (2) $\predicate$ is in $\wedge$-minimal DNF.
\end{customthm}

\begin{proof}
We prove the theorem by induction on the structure of the predicate.  
If $\psi$ is not satisfiable, we have by by Definition~\ref{def:distance} and Algorithm~\ref{alg:pred-val}that $d(v,\psi) = \vpd(v,\psi) = \ozero$. 
Hence, from now on, we consider only satisfiable $\psi$. 
We first prove the theorem for an arbitrary 
term $\psi^{c}$, and then prove it for general DNF formulas $\psi$.
We start with the proof for $\psi^{c}$. 

\noindent \textbf{Base cases:} We have $3$ base cases to consider -
$\psi^{l} \equiv \top$, $\psi^{l} \equiv y \preceq k$ and $\psi^{l} \equiv y \succeq k$.

\begin{itemize}

\item \emph{Case} $\psi^{l} \equiv \top$:

$$
\begin{array}{rcll}
\vpd(v, \top) 	& = 	& \oone & \textrm{by Algorithm~\ref{alg:pred-val}} \\
d(v, \top)		& = 	& \oplus_{v' \models \top} \otimes_{x \in X} d(v(x), v'(x)) & \textrm{by Definition~\ref{def:pred-val}}. \\
			& = 	& \otimes_{x \in X} d(v(x), v(x)) \oplus \oplus_{v' \neq v} \otimes_{x \in X} d(v(x), v'(x)) & \textrm{by semantics of } \top \\
			& = 	& \oone \oplus \oplus_{v' \neq v} \otimes_{x \in X} d(v(x), v'(x)) & \textrm{by Definition~\ref{def:distance}} \\
			& =	& \oone & \textrm{by Definition~\ref{def:semiring}}.
\end{array}
$$

\item \emph{Case} $\psi^{l} \equiv y \preceq k$: We consider two sub-cases, $v(y) \preceq k$ and $v(y) \succ k$.
\begin{itemize}

\item \emph{Case} $v(y) \preceq k$:

$$
\begin{array}{rcll}
\vpd(v, y \preceq k) 	& = 	& \oone & \textrm{by Algorithm~\ref{alg:pred-val}} \\
d(v, y \preceq k)		& = 	& \oplus_{v' \models y \preceq k} \otimes_{x \in X} d(v(x), v'(x)) & \textrm{by Definition~\ref{def:pred-val}} \\
				& = 	& \otimes_{x \in X} d(v(x), v(x)) \oplus &  \\
				& 	& \oplus_{v' \models y \preceq k, v'\neq v}  \otimes_{x \in X} d(v(x), v'(x)) &  \textrm{by associativity of } \oplus \\
				& = 	& \oone &  \\
				& 	& \oplus_{v' \models y \preceq k, v'\neq v''}  \otimes_{x \in X} d(v(x), v'(x)) &  \textrm{by Definition~\ref{def:distance} } \oplus \\
				& = 	& \oone & \textrm{by boundedness of } S. \\
\end{array}
$$

\item \emph{Case} $v(y) \succ k$: Consider arbitrary $a \preceq k$ and $b \succ k$, and arbitrary valuations $v$ such that $v(y) = b$ and 
$v'$ such that $v'(y) = a$ and for all $x \in X\backslash\{y\}$, $v'(x) = v(x)$. We have that

$$
\begin{array}{rcll}
d(v, v'	)			& = 	& \otimes_{x \in X} d(v(x), v'(x)) & \textrm{by Definition~\ref{def:pred-val}} \\
				& = 	& d(v(y),v'(y)) \otimes \otimes_{x \in X\backslash\{y\}} d(v(x), v'(x)) & \\
				& =	& d(b, a) \otimes \otimes_{x \in X\backslash\{y\}} d(v(x), v'(x)) & \textrm{by assumption} \\
				& =	& d(b, a) \otimes \oone & \textrm{by Definition~\ref{def:distance}}. \\

\end{array}
$$

Consider an arbitrary valuation valuation $v''$ and variable $z \in X\backslash\{,x,y\}$ such that $v''(x) = v'(x)$ for  all $x \in X \backslash \{ z \}$. We have that:

$$
\begin{array}{rcll}
\oone				& \sqsubseteq & d(v(z), v''(z)) & \textrm{by Lemma~\ref{lemma:one-min}} \\
d(v, v'	)			& = 	& \otimes_{x \in X} d(v(x), v'(x)) & \textrm{by Definition~\ref{def:pred-val}} \\
				& = 	& d(v(y),v'(y)) \otimes d(v(z),v''(z)) \otimes_{x \in X\backslash\{y,z\}} d(v(x), v'(x)) & \\
				& =	& d(b, a) \otimes d(v(z),v''(z)) \otimes \otimes_{x \in X\backslash\{y,z\}} d(v(x), v'(x)) & \textrm{by assumption} \\
				& =	& d(b, a) \otimes d(v(z), v''(z)) & \textrm{by Definition~\ref{def:distance}}. \\
\end{array}
$$

Combining the facts that $\oone \sqsubseteq d(v(z), v''(z))$ and $d(b,a) \otimes \oone \sqsubseteq d(b,a) \otimes d(v(z), v''(z))$ (Definition~\ref{def:mono} and 
Lemma~\ref{lemma:mono}), we conclude that $d(v,v') \sqsubseteq d(v,v'')$. By Definition~\ref{def:order}, it follows that $d(v,v') \oplus d(v,v'') = d(v,v')$, and hence 
that $\oplus_{v' \models y = a \textrm{s.t.} \; v'(y) = b} d(v, v') = d(b,a) = d(v'(y), a)$.

Consider two arbitrary $a \preceq k$ and $a' \preceq k$ such that $a < a'$. We have that $\oplus_{v' \models y = a \textrm{s.t.} \; v'(y) = b} d(v, v') = d(v'(y),a)$ and 
$\oplus_{v' \models y = a' \textrm{s.t.} \; v'(y) = b} d(v, v') = d(v'(y),a')$. Following the fact that $d(v'(y),a') \sqsubseteq d(v'(y), a)$, 
we conclude that $d(v, y \preceq k) = d(v(y), k)$. 
By Algorithm~\ref{alg:pred-val}, we have that $\vpd(v, y \preceq k) = d(v(y),k)$, hence $\vpd(v, y \preceq k) = d(v, y \preceq k)$.

\end{itemize}

\item \emph{Case} $\psi^{l} \equiv \neg(x \preceq k)$: Symmetric to $\psi^{l} \equiv x \preceq k$.

\end{itemize}

\noindent \textbf{Inductive hypothesis:} $\predicate^{c} \equiv \predicate^{c}_{1} \wedge \predicate^{l}$.

We have $3$ base cases to consider -
$\psi^{l} \equiv \top$, $\psi^{l} \equiv y \preceq k$ and $\psi^{l} \equiv \neg(y \preceq k)$.

\begin{itemize}

\item \emph{Case} $\psi^{l} \equiv \top$: 

$$
\begin{array}{rcll}
\vpd(v, \psi^{c} \wedge \top) 	& = 	& \vpd(v, \psi^{c}) \otimes \oone & \textrm{by Algorithm~\ref{alg:pred-val}} \\
						& =	& \vpd(v, \psi^{c}) & \\
						& =	& d(v, \psi^{c}) & \textrm{by inductive hypothesis} \\
						& =	& d(v, \psi^{c} \wedge \top) & \textrm{by definition of $\top$} \\						
\end{array}
$$

\item \emph{Case} $\psi^{l} \equiv y \preceq k$: We consider two sub-cases, $v(y) \preceq k$ and $\neg(v(y) \preceq k)$.
\begin{itemize}

\item \emph{Case} $v(y) \preceq k$:

$$
\begin{array}{rcll}
\vpd(v, \psi^{c} \wedge y \preceq k) 	& = 	& \vpd(v, \psi^{c}) \otimes \oone & \textrm{by Algorithm~\ref{alg:pred-val}} \\
						& =	& \vpd(v, \psi^{c}) & \\
						& =	& d(v, \psi^{c}) & \textrm{by inductive hypothesis} \\						
\end{array}
$$

Consider an arbitrary $a \preceq k$ and $b \succ k$. Consider two arbitrary valuations $v'$ and $v''$ such that $v'(y) = a$, $v''(y) = b$ and 
for all $x \in X \backslash \{ y \}$, $v'(x) = v''(y)$. We have that $d(v'(y), y \preceq k) = \oone$ and $d(v''(y), y \preceq k) = d(v''(y), k)$ by the 
proof of the base cases, hence $d(v'(y), y \preceq k) \sqsubseteq d(v''(y), y \preceq k)$. By Definition~\ref{def:mono}, we have that 
$d(v, \psi^{c} \wedge y \preceq k) \sqsubseteq d(v, \psi^{c} \wedge \neg(y \preceq k))$. We also have that:

$$
\begin{array}{rcll}
d(v, \psi^{c}) 	& = 	& d(v, \psi^{c} \wedge y \preceq k) \oplus d(v, \psi^{c} \wedge \neg(y \preceq k)) & \textrm{by simple rewriting} \\
			& =	& d(v, \psi^{c} \wedge y \preceq k) & \textrm{by } d(v, \psi^{c} \wedge y \preceq k) \sqsubseteq d(v, \psi^{c} \wedge \neg(y \preceq k)) \\	
						
\end{array}
$$
 
\item \emph{Case} $\neg(v(y) \preceq k)$: We consider two cases, when $S$ is multiplicatively idempotent and when $\psi^{c} \wedge y \preceq k$ is in $\wedge$-minimal 
DNF form.
\begin{itemize}
\item \emph{Case} $S$ \emph{is multiplicatively idempotent}: We first recall that $d(v, y \preceq k) = d(v, y \preceq k) \oplus d(v, \psi^{c} \wedge y \preceq k)$, hence 
by Definition~\ref{def:mono}, $d(y \preceq k) \sqsubseteq  d(v, \psi^{c} \wedge y \preceq k)$.

$$
\begin{array}{lcll}
\vpd(v, \predicate^{c} \wedge y \preceq k) 	& = & \vpd(v, \predicate^{c}) \otimes \vpd(v, y \preceq k) & \textrm{by Algorithm 1} \\
			& = & d(v, \predicate^{c}) \otimes d(v, y \preceq k) & \textrm{by inductive hypothesis} \\
			& = & (d(v, \predicate^{c} \wedge y \preceq k) & \\
			&     & \oplus d(v, \predicate^{c} \wedge \neg(y \preceq k))) & \\
			&     & \otimes d(v, y \preceq k) & \\
			& = & d(v, \predicate^{c} \wedge y \preceq k) \otimes   d(v, y \preceq k) & \textrm{by assumption that } \neg (y \preceq k) \\
			& = & d(v, \predicate^{c} \wedge y \preceq k)& \textrm{by } \otimes \textrm{- idempotence of } S \\
			&&& \textrm{and Definitions~\ref{def:order} and ~\ref{def:mono}} \\
\end{array}
$$

\item \emph{Case} $\psi^{c}$ is in $\wedge$-minimal DNF: By this assumption, $\psi^{c}$ does not contain any conjunct in the form $y \preceq k'$, although it 
might contain a conjunct of the form $\neg(y \preceq k')$ for some $k' \leq k$. However, even if that is the case, by assumption that $v(y) \succ k$, we have that 
the contribution of $y$ in $d(v, \psi^{c})$ is $\oone$. Hence, we have that $d(v, \psi^{c} \wedge y \preceq k)$ consists of computing $d(v, \psi^{c} \wedge)$ and 
$\otimes$-multiplying it with the effect of the $y \preceq k$ constraint, that is with $d(v, y \preceq k) = d(v(y), k)$.

\end{itemize}
\end{itemize}
\item \emph{Case} $\psi^{l} \equiv \neg(y \preceq k):$ symmetric to the previous case.

\end{itemize}

We are now ready to prove that $\vpd(v, \psi \vee \psi^{c}) = d (v, \psi \vee \psi^{c})$.

\noindent \textbf{Base case:} The first part of the proof establishes that $\vpd(v, \psi^{c}) = d (v, \psi^{c})$

\noindent \textbf{Inductive hypothesis:} 
$$
\begin{array}{lcll}
\vpd(v, \predicate) 	& = & \vpd(v, \predicate) \oplus \vpd(v, \predicate^{c}) & \textrm{by Algorithm 1} \\
			& = & d(v, \predicate) \oplus d(v, \psi^{c}) & \textrm{by inductive hypothesis} \\
			& = & \oplus_{v' \models \predicate} \otimes_{x \in X} d(v, v') \; \oplus & \\
			&    & \oplus_{v' \models \predicate^{c}} \otimes_{x \in X} d(v, v') & \textrm{by Definition 5}\\ 
			& = & \oplus_{v' \models \predicate \wedge \neg \predicate^{c}} \otimes_{x \in X} d(v, v') \; \oplus & \\
			&    & \oplus_{v' \models \predicate \wedge \predicate^{c}} \otimes_{x \in X} d(v, v') \; \oplus & \\
			&    & \oplus_{v' \models \predicate \wedge \predicate^{c}} \otimes_{x \in X} d(v, v') \; \oplus & \\
			&    & \oplus_{v' \models \neg \predicate \wedge \predicate^{c}} \otimes_{x \in X} d(v, v') & \textrm{by partition of sets} \\
			& = & \oplus_{v' \models \predicate \wedge \neg \predicate^{c}} \otimes_{x \in X} d(v, v') \; \oplus & \\
			&    & \oplus_{v' \models \predicate \wedge \predicate^{c}} \otimes_{x \in X} d(v, v') \; \oplus & \\
			&    & \oplus_{v' \models \neg \predicate \wedge \predicate^{c}} \otimes_{x \in X} d(v, v') & \textrm{by idempotence of } \oplus \\
			& = & \oplus_{v' \models \predicate \vee \predicate^{c}} \otimes_{x \in X} d(v, v') & \textrm{by union of disjoint sets} \\
			& = & d(v, \predicate \vee \predicate^{c}) & \textrm{by Definition 5}.
\end{array}
$$

\end{proof} \qed

\begin{customthm}{\ref{th:val}}
Given a specification $\varphi$, its associated SWA $\waut$ defined over a semiring $S$ and a trace $\trace$, we have that 
$\textsf{val}(\trace, \waut) = \tvalue(\trace, \waut) = d(\trace, \varphi)$.
\end{customthm}

\begin{proof}
The proof follows from the monotonicity of the natural order in additively idempotent semirings. This propery allow us 
to merge the values of all paths $\pi$ induced by a prefix of size $n$ of the trace $ \tau$ and ending in a location $q$ 
into a single representative value that we represent as the cost $c(q,n)$ of the location at time $n$ and that is used to 
compute the value of all the extensions.  (we recall that 
for all $a,b,c \in S$, if $a \sqsubseteq b$, then $a \otimes c \sqsubseteq b \otimes c$).  Following the definition of the 
trace value, the cost $c(q,n+1)$ of $q$ at time $n+1$ is then the $\oplus$-summation of the individual effects 
of taking all possible transitions $(s,\psi, q)$ from $s$ to $q$ with the new valuation $v_{n+1}$ (the cost $c(s,n)$ 
$\otimes$-multiplied by the predicate-value distance $\vpd(v_{n+1}, \psi)$).

\end{proof} \qed

\begin{customthm}{\ref{th:qual}}
Given traces $\trace$ and $\trace'$, a specification $\varphi$ and distances $d(\tau, \tau')$, $d(\tau, \neg \varphi)$ defined over a bounded semiring $S$, 
$$
\begin{array}{rcl}
\rho(\trace, \varphi) > 0 & \rightarrow & \tau \models \varphi \\
\rho(\trace, \varphi) < 0 & \rightarrow & \tau \not \models \varphi \\
\trace \models \spec \textrm{ and } d(\trace, \trace') \sqsubset d(\tau, \neg \varphi) & \rightarrow & \trace' \models \spec.
\end{array}
$$
\end{customthm}

\begin{proof}
The proof for the first two implications is trivial from the definitions of $\rho(\trace, \varphi)$ and $d(\trace, \varphi)$. We prove the third 
implication by contradiction. Assume that $\trace \models \spec$, $d(\trace, \trace') \sqsubset d(\trace, \neg \spec)$ and $\tau' \not \models \spec$. 
Then, by definition of the distance, we have that $d(\trace', \neg \varphi) = \oone$. 
By the additive idempotence of $S$ and the definition of $d(\trace', \neg \varphi)$, 
there exists $\tau'' \models \neg \varphi$ such that $d(\trace', \trace'') = \oone$, hence 
$\trace' = \trace''$.  However, in that case we have that $d(\tau, \tau') = d(\tau, \neg \varphi)$, which is a contradiction.
\end{proof} \qed

%% file: main-with-authos.bbl
\begin{thebibliography}{10}

\bibitem{fals2}
Houssam Abbas, Bardh Hoxha, Georgios~E. Fainekos, Jyotirmoy~V. Deshmukh, James
  Kapinski, and Koichi Ueda.
\newblock Conformance testing as falsification for cyber-physical systems.
\newblock {\em CoRR}, abs/1401.5200, 2014.

\bibitem{AbbasMF14}
Houssam Abbas, Hans~D. Mittelmann, and Georgios~E. Fainekos.
\newblock Formal property verification in a conformance testing framework.
\newblock In {\em Proc. of {MEMOCODE} 2014: the Twelfth {ACM/IEEE}
  International Conference on Formal Methods and Models for Codesign}, pages
  155--164. {IEEE}, 2014.

\bibitem{avstl}
Takumi Akazaki and Ichiro Hasuo.
\newblock Time robustness in {MTL} and expressivity in hybrid system
  falsification.
\newblock In {\em Computer Aided Verification - 27th International Conference,
  {CAV} 2015, San Francisco, CA, USA, July 18-24, 2015, Proceedings, Part
  {II}}, pages 356--374, 2015.

\bibitem{staliro_base}
Yashwanth Annpureddy, Che Liu, Georgios~E. Fainekos, and Sriram
  Sankaranarayanan.
\newblock {S}-{T}a{L}i{R}o: {A} tool for temporal logic falsification for
  hybrid systems.
\newblock In {\em Proc. of {TACAS} 2011: the 17th International Conference on
  Tools and Algorithms for the Construction and Analysis of Systems}, volume
  6605 of {\em LNCS}, pages 254--257. Springer, 2011.

\bibitem{OdedSRE}
Alexey Bakhirkin, Thomas Ferr{\`{e}}re, Oded Maler, and Dogan Ulus.
\newblock On the quantitative semantics of regular expressions over real-valued
  signals.
\newblock In {\em Formal Modeling and Analysis of Timed Systems - 15th
  International Conference, {FORMATS} 2017, Berlin, Germany, September 5-7,
  2017, Proceedings}, pages 189--206, 2017.

\bibitem{BartocciBLN17}
Ezio Bartocci, Luca Bortolussi, Michele Loreti, and Laura Nenzi.
\newblock Monitoring mobile and spatially distributed cyber-physical systems.
\newblock In {\em Proc.of {MEMOCODE} 2017: the 15th {ACM-IEEE} International
  Conference on Formal Methods and Models for System Design}, pages 146--155.
  {ACM}, 2017.

\bibitem{chapter5}
Ezio Bartocci, Jyotirmoy Deshmukh, Alexandre Donz\'{e}, Georgios Fainekos, Oded
  Maler, Dejan Nickovic, and Sriram Sankaranarayanan.
\newblock Specification-based monitoring of cyber-physical systems: A survey on
  theory, tools and applications.
\newblock In {\em Lectures on Runtime Verification - Introductory and Advanced
  Topics}, volume 10457 of {\em LNCS}, pages 128--168. Springer, 2018.

\bibitem{bellman1958routing}
Richard Bellman.
\newblock On a routing problem.
\newblock {\em Quarterly of applied mathematics}, 16(1):87--90, 1958.

\bibitem{stlstarq}
Lubos Brim, Tomas Vejpustek, David Safr{\'{a}}nek, and Jana Fabrikov{\'{a}}.
\newblock Robustness analysis for value-freezing signal temporal logic.
\newblock In {\em Proceedings Second International Workshop on Hybrid Systems
  and Biology, {HSB} 2013, Taormina, Italy, 2nd September 2013.}, pages 20--36,
  2013.

\bibitem{chatterjee2008quantitative}
Krishnendu Chatterjee, Laurent Doyen, and Thomas~A Henzinger.
\newblock Quantitative languages.
\newblock In {\em International Workshop on Computer Science Logic}, pages
  385--400. Springer, 2008.

\bibitem{LOLA}
Ben D'Angelo, Sriram Sankaranarayanan, C{\'{e}}sar S{\'{a}}nchez, Will
  Robinson, Bernd Finkbeiner, Henny~B. Sipma, Sandeep Mehrotra, and Zohar
  Manna.
\newblock {LOLA:} runtime monitoring of synchronous systems.
\newblock In {\em 12th International Symposium on Temporal Representation and
  Reasoning {(TIME} 2005), 23-25 June 2005, Burlington, Vermont, {USA}}, pages
  166--174, 2005.

\bibitem{simmin}
Loris D'Antoni and Margus Veanes.
\newblock Minimization of symbolic automata.
\newblock In {\em The 41st Annual {ACM} {SIGPLAN-SIGACT} Symposium on
  Principles of Programming Languages, {POPL} '14, San Diego, CA, USA, January
  20-21, 2014}, pages 541--554, 2014.

\bibitem{loris2}
Loris D'Antoni and Margus Veanes.
\newblock Extended symbolic finite automata and transducers.
\newblock {\em Formal Methods in System Design}, 47(1):93--119, 2015.

\bibitem{loris1}
Loris D'Antoni and Margus Veanes.
\newblock The power of symbolic automata and transducers.
\newblock In {\em Computer Aided Verification - 29th International Conference,
  {CAV} 2017, Heidelberg, Germany, July 24-28, 2017, Proceedings, Part {I}},
  pages 47--67, 2017.

\bibitem{de2008z3}
Leonardo De~Moura and Nikolaj Bj{\o}rner.
\newblock Z3: An efficient smt solver.
\newblock {\em Tools and Algorithms for the Construction and Analysis of
  Systems}, pages 337--340, 2008.

\bibitem{DeshmukhDGJJS17}
Jyotirmoy~V. Deshmukh, Alexandre Donz{\'{e}}, Shromona Ghosh, Xiaoqing Jin,
  Garvit Juniwal, and Sanjit~A. Seshia.
\newblock Robust online monitoring of signal temporal logic.
\newblock {\em Formal Methods in System Design}, 51(1):5--30, 2017.

\bibitem{DokhanchiHF14}
Adel Dokhanchi, Bardh Hoxha, and Georgios~E. Fainekos.
\newblock On-line monitoring for temporal logic robustness.
\newblock In {\em Proc. {RV} 2014: the 5th International Conference on Runtime
  Verification}, volume 8734 of {\em Lecture Notes in Computer Science}, pages
  231--246. Springer, 2014.

\bibitem{fals1}
Adel Dokhanchi, Aditya Zutshi, Rahul~T. Sriniva, Sriram Sankaranarayanan, and
  Georgios~E. Fainekos.
\newblock Requirements driven falsification with coverage metrics.
\newblock In {\em 2015 International Conference on Embedded Software, {EMSOFT}
  2015, Amsterdam, Netherlands, October 4-9, 2015}, pages 31--40, 2015.

\bibitem{breach}
Alexandre Donz{\'{e}}.
\newblock Breach, {A} toolbox for verification and parameter synthesis of
  hybrid systems.
\newblock In {\em Computer Aided Verification, 22nd International Conference,
  {CAV} 2010, Edinburgh, UK, July 15-19, 2010. Proceedings}, pages 167--170,
  2010.

\bibitem{robust2}
Alexandre Donz{\'{e}}, Thomas Ferr{\`{e}}re, and Oded Maler.
\newblock Efficient robust monitoring for {STL}.
\newblock In {\em Computer Aided Verification {(CAV)}}, pages 264--279, 2013.

\bibitem{robust1}
Alexandre Donz{\'{e}} and Oded Maler.
\newblock Robust satisfaction of temporal logic over real-valued signals.
\newblock In {\em Formal Modeling and Analysis of Timed Systems {(FORMATS)}},
  pages 92--106, 2010.

\bibitem{stltest}
Tommaso Dreossi, Thao Dang, Alexandre Donz{\'{e}}, James Kapinski, Xiaoqing
  Jin, and Jyotirmoy~V. Deshmukh.
\newblock Efficient guiding strategies for testing of temporal properties of
  hybrid systems.
\newblock In {\em {NASA} Formal Methods - 7th International Symposium, {NFM}
  2015, Pasadena, CA, USA, April 27-29, 2015, Proceedings}, pages 127--142,
  2015.

\bibitem{fainekos-robust}
Georgios~E. Fainekos and George~J. Pappas.
\newblock Robustness of temporal logic specifications for continuous-time
  signals.
\newblock {\em Theor. Comput. Sci.}, 410(42):4262--4291, 2009.

\bibitem{njofra}
Georgios~E. Fainekos, Sriram Sankaranarayanan, Franjo Ivancic, and Aarti Gupta.
\newblock Robustness of model-based simulations.
\newblock In {\em Proc. of {RTSS} 2009: the 30th {IEEE} Real-Time Systems
  Symposium}, pages 345--354. {IEEE} Computer Society, 2009.

\bibitem{measures}
Thomas Ferr\`{e}re, Oded Maler, Dejan Nickovic, and Dogan Ulus.
\newblock Measuring with timed patterns.
\newblock In {\em Computer Aided Verification, 27th International Conference,
  {CAV} 2015, San Francisco, CA, USA, July 18-24, 2011, Proceedings}, 2015.

\bibitem{ltl-tableau}
Rob Gerth, Doron Peled, Moshe~Y Vardi, and Pierre Wolper.
\newblock Simple on-the-fly automatic verification of linear temporal logic.
\newblock In {\em Protocol Specification, Testing and Verification XV}, pages
  3--18. Springer, 1995.

\bibitem{hoxha_bench}
Bardh Hoxha, Houssam Abbas, and Georgios~E. Fainekos.
\newblock Benchmarks for temporal logic requirements for automotive systems.
\newblock In {\em 1st and 2nd International Workshop on Applied veRification
  for Continuous and Hybrid Systems, ARCH@CPSWeek 2014, Berlin, Germany, April
  14, 2014 / ARCH@CPSWeek 2015, Seattle, WA, USA, April 13, 2015.}, pages
  25--30, 2014.

\bibitem{stledit}
Stefan Jaksic, Ezio Bartocci, Radu Grosu, and Dejan Nickovic.
\newblock Quantitative monitoring of {STL} with edit distance.
\newblock In {\em Runtime Verification - 16th International Conference, {RV}
  2016, Madrid, Spain, September 23-30, 2016, Proceedings}, pages 201--218,
  2016.

\bibitem{koymans}
Ron Koymans.
\newblock Specifying real-time properties with metric temporal logic.
\newblock {\em Real-Time Systems}, 2(4):255--299, 1990.

\bibitem{mn04}
Oded Maler and Dejan Nickovic.
\newblock Monitoring temporal properties of continuous signals.
\newblock In {\em Formal Techniques, Modelling and Analysis of Timed and
  Fault-Tolerant Systems, Joint International Conferences on Formal Modelling
  and Analysis of Timed Systems, {FORMATS} 2004 and Formal Techniques in
  Real-Time and Fault-Tolerant Systems, {FTRTFT} 2004, Grenoble, France,
  September 22-24, 2004, Proceedings}, pages 152--166, 2004.

\bibitem{mn13}
Oded Maler and Dejan Nickovic.
\newblock Monitoring properties of analog and mixed-signal circuits.
\newblock {\em {STTT}}, 15(3):247--268, 2013.

\bibitem{mohri_semiring}
Mehryar Mohri.
\newblock Semiring frameworks and algorithms for shortest-distance problems.
\newblock {\em Journal of Automata, Languages and Combinatorics},
  7(3):321--350, 2002.

\bibitem{edit-weighted}
Mehryar Mohri.
\newblock Edit-distance of weighted automata: General definitions and
  algorithms.
\newblock {\em Int. J. Found. Comput. Sci.}, 14(6):957--982, 2003.

\bibitem{amt}
Dejan Nickovic and Oded Maler.
\newblock {AMT:} {A} property-based monitoring tool for analog systems.
\newblock In {\em Formal Modeling and Analysis of Timed Systems, 5th
  International Conference, {FORMATS} 2007, Salzburg, Austria, October 3-5,
  2007, Proceedings}, pages 304--319, 2007.

\bibitem{testers}
Amir Pnueli and Aleksandr Zaks.
\newblock On the merits of temporal testers.
\newblock In {\em 25 Years of Model Checking - History, Achievements,
  Perspectives}, pages 172--195, 2008.

\bibitem{stlce}
Vasumathi Raman, Alexandre Donz{\'{e}}, Dorsa Sadigh, Richard~M. Murray, and
  Sanjit~A. Seshia.
\newblock Reactive synthesis from signal temporal logic specifications.
\newblock In {\em Proceedings of the 18th International Conference on Hybrid
  Systems: Computation and Control, HSCC'15, Seattle, WA, USA, April 14-16,
  2015}, pages 239--248, 2015.

\bibitem{rizk}
Aur{\'{e}}lien Rizk, Gr{\'{e}}gory Batt, Fran{\c{c}}ois Fages, and Sylvain
  Soliman.
\newblock On a continuous degree of satisfaction of temporal logic formulae
  with applications to systems biology.
\newblock In {\em Computational Methods in Systems Biology, 6th International
  Conference, {CMSB} 2008, Rostock, Germany, October 12-15, 2008. Proceedings},
  pages 251--268, 2008.

\bibitem{rodionova2016temporal}
Alena Rodionova, Ezio Bartocci, Dejan Nickovic, and Radu Grosu.
\newblock Temporal logic as filtering.
\newblock In {\em Proceedings of the 19th International Conference on Hybrid
  Systems: Computation and Control}, pages 11--20. ACM, 2016.

\bibitem{alena_bench}
Alena Rodionova, Matthew O'Kelly, Houssam Abbas, Vincent Pacelli, and Rahul
  Mangharam.
\newblock An autonomous vehicle control stack.
\newblock In {\em {ARCH17.} 4th International Workshop on Applied Verification
  of Continuous and Hybrid Systems, collocated with Cyber-Physical Systems Week
  (CPSWeek) on April 17, 2017 in Pittsburgh, PA, {USA}}, pages 44--51, 2017.

\bibitem{roopsha_robustness13}
Roopsha Samanta, Jyotirmoy~V. Deshmukh, and Swarat Chaudhuri.
\newblock Robustness analysis of string transducers.
\newblock In {\em Proc. of {ATVA} 2013: the 11th International Symposium on
  Automated Technology for Verification and Analysis}, volume 8172 of {\em
  LNCS}, pages 427--441. Springer, 2013.

\bibitem{veanes_symbolic}
Margus Veanes, Pieter Hooimeijer, Benjamin Livshits, David Molnar, and Nikolaj
  Bj{\o}rner.
\newblock Symbolic finite state transducers: algorithms and applications.
\newblock In {\em Proceedings of the 39th {ACM} {SIGPLAN-SIGACT} Symposium on
  Principles of Programming Languages, {POPL} 2012, Philadelphia, Pennsylvania,
  USA, January 22-28, 2012}, pages 137--150, 2012.

\end{thebibliography}
